\documentclass[aps,prb,twocolumn,amssymb,floatfix,epsfig,showlabel]{revtex4}
\usepackage{graphicx,epsfig,psfrag,subfigure,amsopn,epstopdf}
\usepackage{color,ulem,verbatim,float,wrapfig}
\usepackage{hyperref}
\usepackage{hypcap}
\usepackage{amssymb,amsbsy,amsmath,mathrsfs}
\usepackage{color}
\usepackage{ulem}
\usepackage{setspace}
\usepackage{float}
\definecolor{green}{rgb}{0.0,0.4,0}
\newcommand\bib{\bibitem}
\newcommand\mc{\mathcal}
\newcommand\beq{\begin{equation}}
\newcommand\eeq{\end{equation}}
\newcommand\bes{\begin{subequations}}
\newcommand\ees{\end{subequations}}
\newcommand\bea{\begin{eqnarray}}
\newcommand\eea{\end{eqnarray}}
\newcommand\sg{\subfigure}
\newcommand\non{\nonumber}

\newcommand\ig{\includegraphics}

\newcommand\ga{\gamma}

\newcommand\De{\Delta}
\newcommand\ep{\epsilon}
\newcommand\si{\sigma}

\newcommand\la{\langle}
\newcommand\ra{\rangle}

\newcommand\ua{\uparrow}
\newcommand\da{\downarrow}
\newcommand{\bk}{{\bf k}}

\newcommand{\bs}{\bar{s}}

\begin{document}

\title{Generating a second-order topological insulator with multiple 
corner states by periodic driving}
\author{Ranjani Seshadri, Anirban Dutta and Diptiman Sen}
\affiliation{\small{Centre for High Energy Physics, Indian Institute of
Science, Bengaluru 560012, India}}

\begin{abstract}
We study the effects of periodic driving on a variant of the 
Bernevig-Hughes-Zhang (BHZ) model defined on a square lattice. In the absence 
of driving, the model has both topological and nontopological phases 
depending on the different parameter values. We also study the anisotropic 
BHZ model and show that, unlike the isotropic model, it has a nontopological 
phase which has states localized on only two of the four edges of a 
finite-sized square. When an appropriate term is added, the 
edge states get gapped and gapless states appear at the four corners 
of a square; we have shown that these corner states can be labeled by the 
eigenvalues of a certain operator. When the system is driven periodically by a 
sequence of two pulses, multiple corner states may appear depending on the 
driving frequency and other parameters. We discuss to what extent the system 
can be characterized by topological invariants such as the Chern number and 
a diagonal winding number. We have shown that the locations of the jumps in 
these invariants can be understood in terms of the Floquet operator at both 
the time-reversal invariant momenta and other momenta which have no 
special symmetries.
\end{abstract}

\date{\today}

\maketitle

\section{Introduction}
\label{sec:intro}

Topological insulators (TIs) have been studied extensively for the last 
several years~\cite{hasan,qi}. A key feature of these materials is that the 
bulk states are gapped but there are gapless states at the boundaries which 
contribute to transport and other properties at low temperatures. Further, 
there is a bulk-boundary correspondence, namely, the bulk bands are 
characterized by a topological invariant (such as the Chern number for 
two-dimensional TIs~\cite{hasan,qi}) which is an integer, and the number 
of states with a particular momentum at each of the boundaries is equal to 
the topological invariant. Recently, a generalization of these materials 
called higher-order TIs has been introduced~\cite{benalcazar}. For instance, 
a second-order TI in two dimensions is a system in which the bulk and edge 
states are both gapped but there are gapless states at the corners of the 
system~\cite{peng1,langbehn,song,ezawa,dwivedi,miert,schindler,franca,xie,
calugaru, kang,queiroz,hsu,matsugatani}. An electrical circuit realization of 
such a system has been reported recently~\cite{imhof}.

Closed quantum systems driven periodically in time constitute another 
area that has been studied by several groups in recent years~\cite{dziarmaga,
polkovnikov,dutta,dalessio}. In particular, there has been much interest in 
understanding the conditions under which periodic driving can generate 
topological phases and boundary modes~\cite{top1,top2,top3,top4,top5,top6,
top12,top15,top16,top17,top18,top19,top21,top22,top24,top25,top26}. It is 
sometimes found that even if the time-dependent
Hamiltonian lies in a nontopological phase at each instant of time, the 
unitary time-evolution operator for one time period (called the Floquet 
operator) has eigenstates which are localized near the boundaries of the 
system. It is therefore interesting to investigate whether higher-order TIs 
can also be generated by periodic driving, for example, whether such a 
driving can generate corner states in a two-dimensional 
system which has no such states in the absence of driving. 

There have been some earlier studies of the generation of higher-order 
TIs by periodic driving~\cite{bomantara,huang,peng2,rodriguez2}. Using a 
mirror symmetry which is present in these models, a mirror-graded winding 
number was derived which can predict the number of corner or hinge states 
which appear as a result of the driving.

In contrast to the above studies, we will study a variant of the 
well-known Bernevig-Hughes-Zhang (BHZ) model which, we will show, hosts corner 
states for certain values of the system parameters. We will then study what 
happens when one of the parameters is varied periodically in time and show 
that this can generate corner states. We will then consider an anisotropic 
version of the BHZ model which has not been studied before to the best
of our knowledge. Unlike the isotropic model,
this has a nontopological phase in which only two out of the four edges of 
a finite-sized square has edge states which are not topologically protected. 
We find that this model can also have corner states, with or without driving. 
Interestingly, we find that driving in certain parameter regimes can generate 
more than one state at each corner, unlike the time-independent model which 
never has more than one localized state at each corner. Following earlier 
papers~\cite{schindler,imhof,top22,asboth,yao,rodriguez,bomantara,huang,peng2,
rodriguez2}, we will also study two topological invariants, a Chern number and 
a winding number, to see whether these can be used to understand the edge and 
corner states which appear in the different phases of our model. Interestingly,
we find that the winding number can be defined for both the isotropic and 
anisotropic models, even though only the isotropic model has a 
mirror symmetry.

The plan of this paper is as follows. In Sec.~\ref{sec:st} we present the 
Hamiltonian of an isotropic model constructed by adding an appropriate term 
to the BHZ Hamiltonian; this additional term is essential to generate corner 
states~\cite{schindler}. After analyzing the different symmetries of the 
Hamiltonian, we will discuss how two topological invariants called the Chern 
number~\cite{hasan,qi} and a `diagonal' winding 
number~\cite{schindler,imhof,rodriguez2} can be used to predict when gapless
edge corner states will appear. We then discuss how to numerically study edge 
states by looking at a ribbon which is infinitely long but has a finite width 
and corner states by looking at a finite-sized square. We have shown that
the corner states can be labeled by the eigenvalues ($\pm 1$) of a certain 
operator. In Sec.~\ref{sec:driven}, we study the effects of periodic driving 
on the isotropic model. To this end, we numerically calculate the Floquet 
operator and find its eigenvalues and eigenstates. We show how this modifies 
the regions of nonzero Chern and diagonal winding numbers; depending on the 
parameters, the magnitudes of the Chern number and diagonal winding number
can be larger than 1. We also 
find that the driving can generate corner states; these states always have 
Floquet eigenvalues equal to $\pm 1$. In Sec.~\ref{sec:st_asym}, we study an 
anisotropic model in which the hopping parameter has different values in the
$x$ and $y$ directions. 
This model has a nontopological phase where the Chern number is zero but the 
winding number is nonzero; in this phase, there are edge states on only two 
of the edges of a finite-sized square which are not topologically protected.
We again study the effects of periodic driving and show that this can generate
corner states. We find that there may be more than one state at each corner
if the driving frequency is low. In Sec.~\ref{sec:disc} we summarize our 
results and point out some directions for future research. Finally, we discuss 
in Appendix A how the locations as well as the magnitudes of the jumps in the 
Chern number and diagonal winding number for the periodically driven system can
be understood in terms of the contributions from the time-reversal invariant 
momenta, and in Appendix B how the jumps in the Chern number can occur 
due to contributions from other momenta where there are no special symmetries.

\section{Time-independent isotropic model}
\label{sec:st}

In this section, we will study the properties of a variant of the BHZ model 
which is motivated by a three-dimensional model discussed in 
Ref.~\onlinecite{schindler}.
The purpose of this is to contrast the properties of this system with that 
of the periodically driven system that we will study in Sec.~\ref{sec:driven}.

\subsection{Bulk Hamiltonian}
\label{sec:st_bulk}

We will consider the following Hamiltonian for a system with periodic boundary
conditions in which the momentum $\bk = (k_x,k_y)$ is a good quantum 
number~\cite{schindler}
\bea H (\bk)&=& \big[ M ~+~ t_0 ~(\cos k_x ~+~ \cos k_y)\big] ~\tau^z 
\otimes \si^0 \non \\
&&+ ~\De_1 ~(\sin k_x ~\tau^x \otimes \si^x ~+~ \sin k_y ~\tau^x \otimes \si^y)
\non \\
&&+ ~\De_2 ~(\cos k_x ~-~\cos k_y) ~\tau^y \otimes \si^0, 
\label{eq:con_ham} \eea
where $\vec \tau$ and $\vec \si$ are Pauli matrices acting on the orbital 
and spin degrees of freedom respectively, and $\tau^0$ and $\si^0$ denote
$2 \times 2$ identity matrices in these two spaces respectively. In 
Eq.~\eqref{eq:con_ham}, $t_0$ denotes a spin-independent but orbital-dependent
hopping amplitude 
between nearest-neighbor sites, and $\De_1$ denotes a spin-orbit coupling.
We will find later that the last term with coefficient $\De_2$ is necessary in
order to have corner states, with or without periodic driving. This 
term corresponds to orbital currents which break time-reversal symmetry 
oppositely in the $x$ and $y$ directions; physical systems where such a 
term can appear are described in Ref.~\onlinecite{schindler}.

We first study the symmetries of the Hamiltonian in Eq.~\eqref{eq:con_ham} 
for the case with $\De_2=0$.

\begin{enumerate}

\item {\it Time-reversal $\mc{T}$}: For $\De_2=0$, we have the 
BHZ model~\cite{bhz}, a well-known example of a 
two-dimensional TI. The Hamiltonian is invariant under the time-reversal 
transformation, i.e., $\mc{T}H (\bk)\mc{T}^{-1} =H (-\bk)$, where $\mc{T}=
\tau^0\si^y \mc{K}$, and $\mc{K}$ denotes the complex conjugation operator. 

\item {\it Charge conjugation $\mc{C}$}: $H (\bk)$ has a particle-hole 
or charge conjugation symmetry given by $\mc{C} = \tau^y \si^y \mc{K}$ such 
that $\mc{C}H (\bk)\mc{C}^{-1} =-H^{*}(-\bk)$, and a chiral symmetry 
$\mc{S}_1 = \mc{CT} = \tau^y \si^0$ with $\mc{S}_1 H (\bk) \mc{S}_1^{-1} = -
H (\bk)$. Therefore, the system belongs to the BDI class of the 
Altland-Zirnbauer classification~\cite{altland}. There is another 
symmetry operator $\mc{S}_2 = \tau^x \si^z$ which gives $\mc{S}_2 H (\bk) 
\mc{S}_2^{-1} = - H (\bk)$.

\item {\it Fourfold rotation $\mc{C}_4$}: $H (\bk)$ has a fourfold rotation 
symmetry about the $z$ axis, i.e., $\mc{C}_4 H (\bk) (\mc{C}_4)^{-1}=H 
(\mc{C}_4 \bk)$, where $\mc{C}_4=\tau^0 e^{-i(\pi/4) \si^z}$ and 
$\mc{C}_4 (k_x,k_y) =(k_y,-k_x)$.

\item The operator $\mc{P}_1= \mc{S}_1 \mc{S}_2 \tau^z\si^z$ commutes with 
the Hamiltonian. We note that $\mc{P}_1^2 = \mathbb{I}_4$, the $4 \times 4$ 
identity matrix. We will show later (see the discussion after 
Eq.~\eqref{spec1}) that there is a twofold degeneracy of the energy spectrum 
for each value of $\bk$; the two degenerate states have eigenvalues of 
$\mc{P}_1 = \pm 1$. 

\end{enumerate}

\begin{widetext}
\begin{figure}[htb]
\begin{center}
\sg[~Band gap]{\ig[height = 3cm]{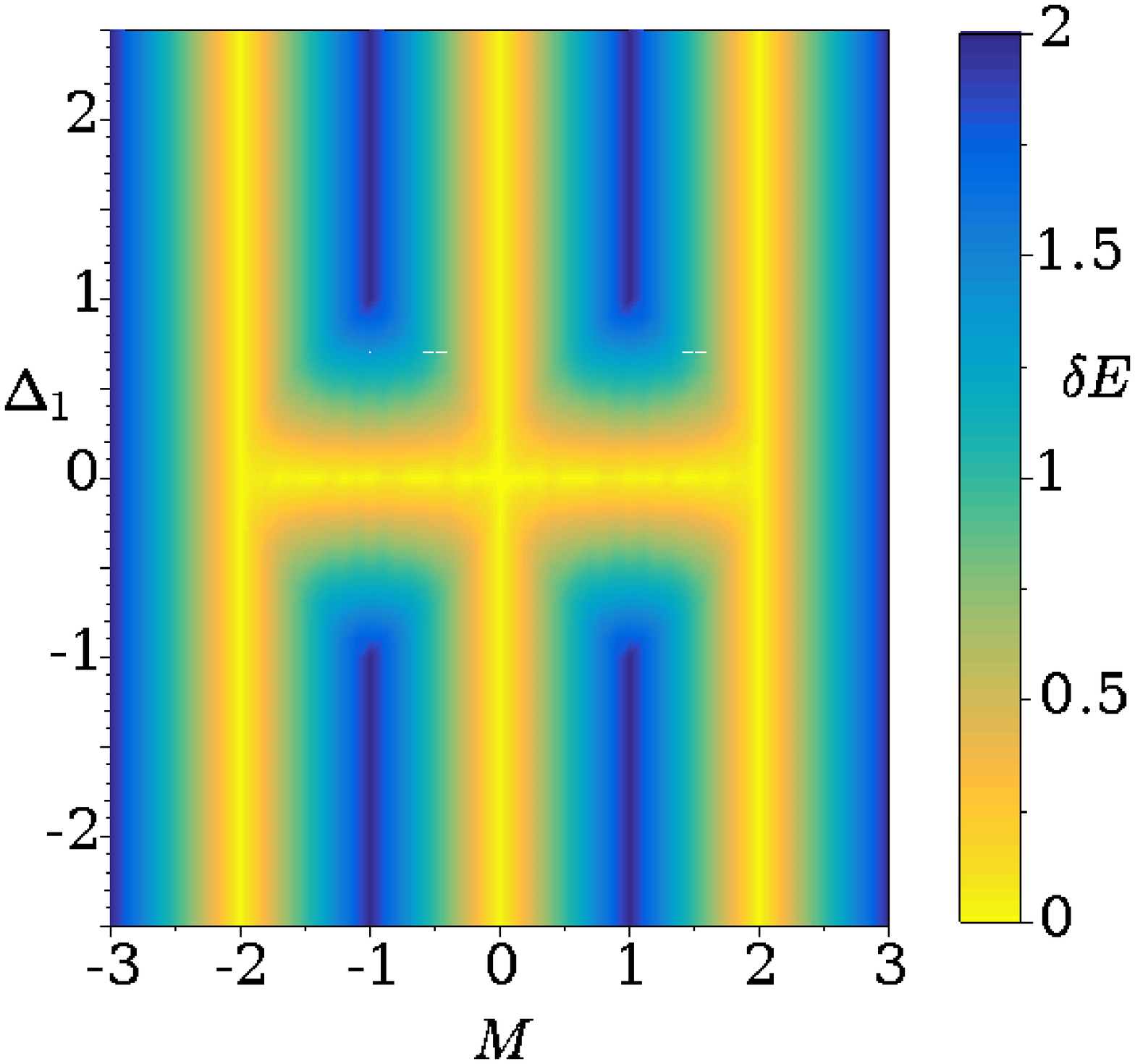}\label{fig:E_gap}} 
\sg[~Chern number]{\ig[height=3cm]{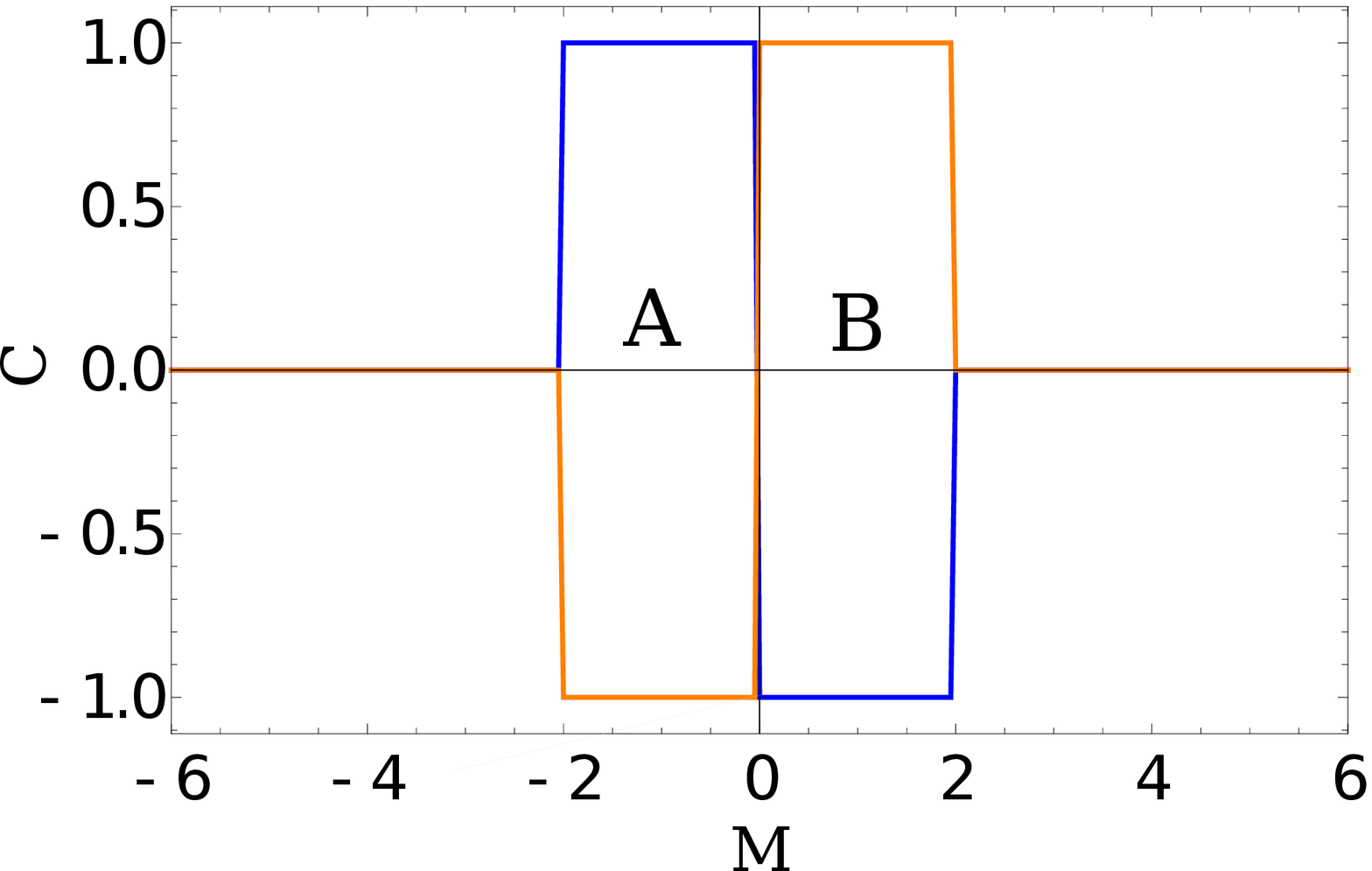}\label{fig:chern_eq}} 
\sg[~Winding number]{\ig[height=3cm]{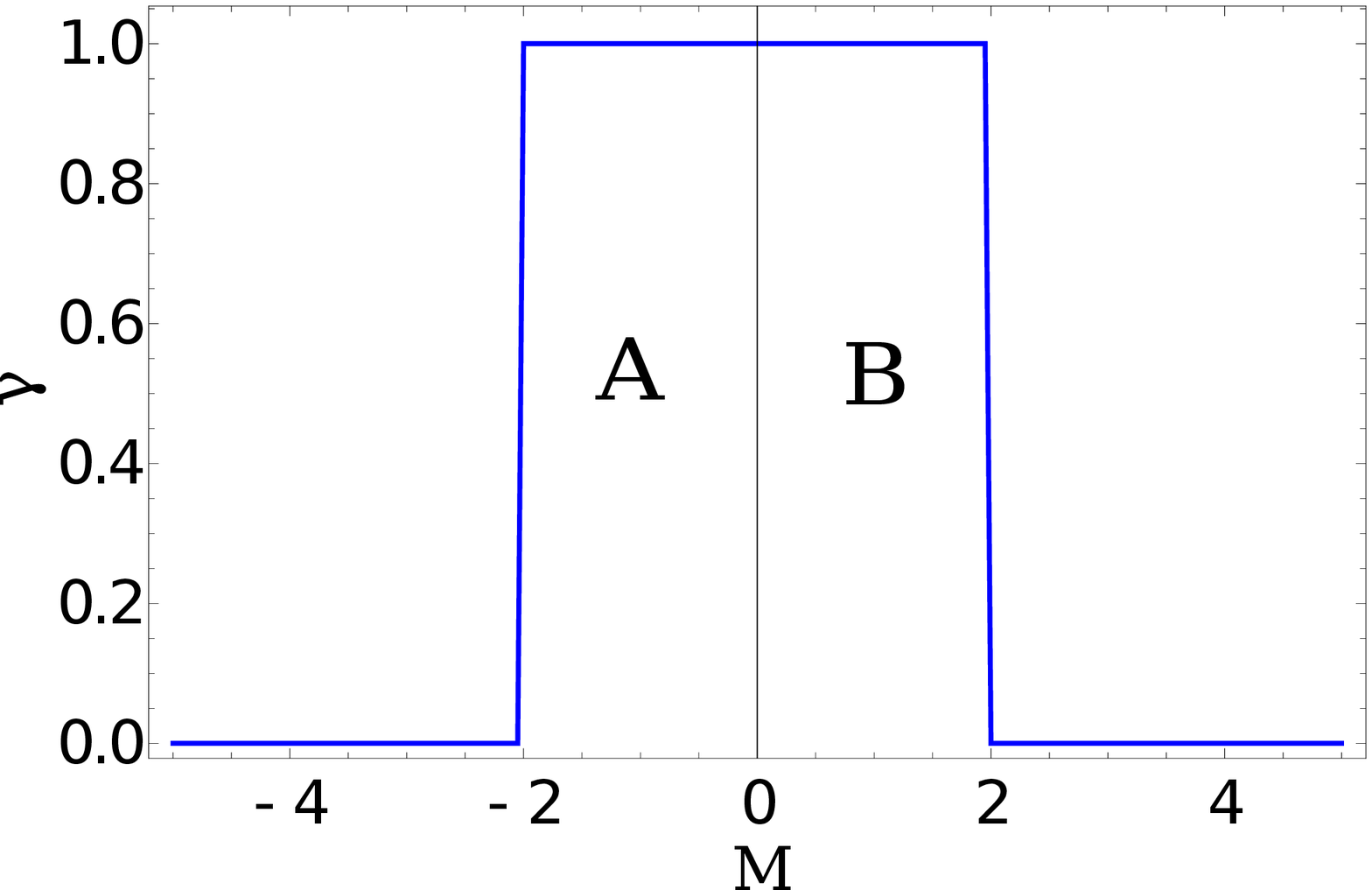}\label{fig:mirror_eq}}
\sg[~Winding number = 1]{\ig[height=3cm]{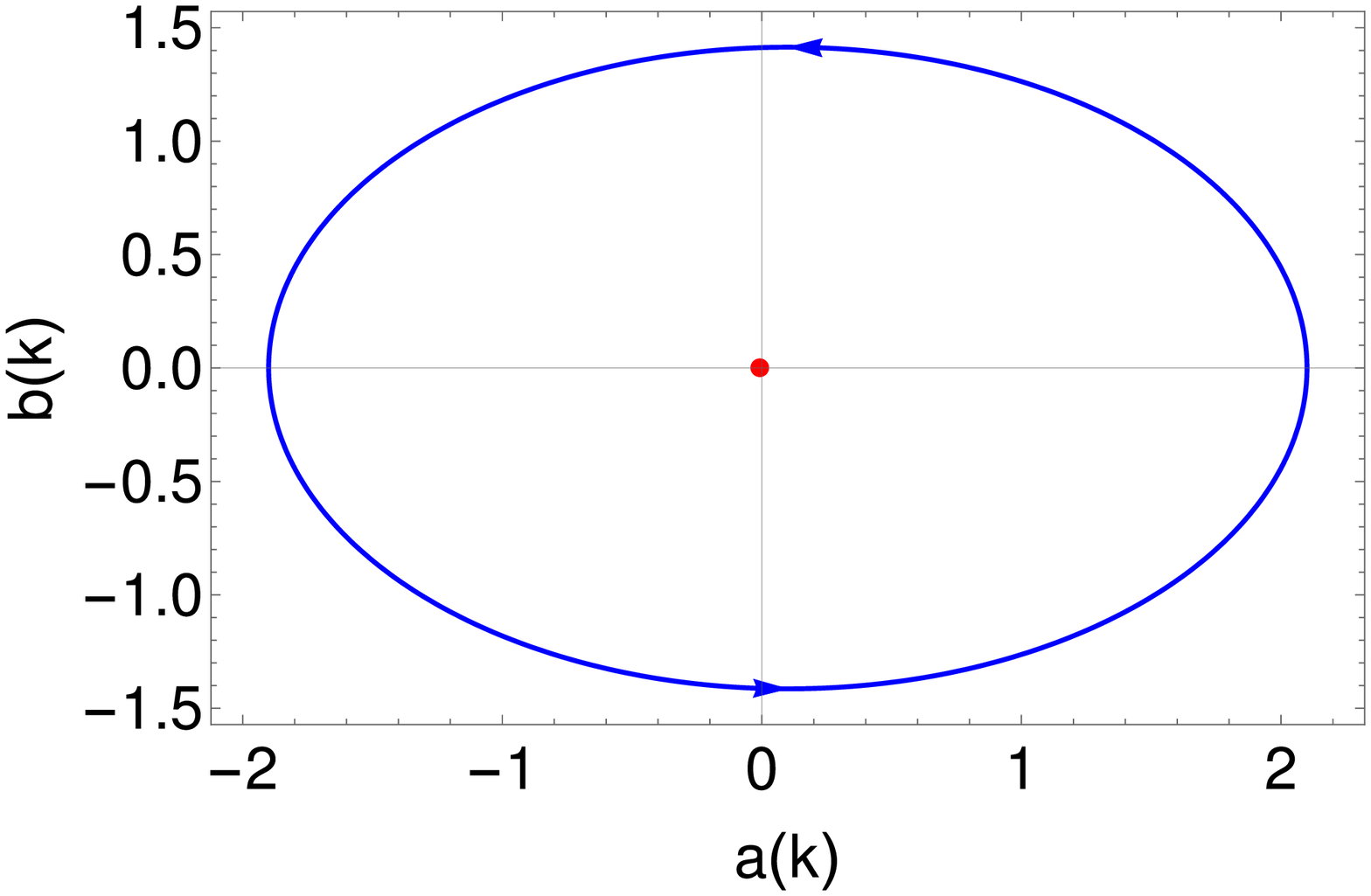}\label{fig:mw1_eq}}
\caption[]{(a) Energy gap as a function of the parameters $M$ and $\De_1$ for 
$\De_2 = 0$. The range of $M$ taken here is the region where the system is 
topological. The band gap is zero along the lines $M = 0, \pm 2$. 
(b) Chern number for the Hamiltonian in Eq.~\eqref{eq:con_ham} with $\De_2=0$. 
The horizontal solid line shows the numerically calculated Chern number $C$
separating the topological $|M| < 2$ ($C \ne 0$) and nontopological regions 
($C = 0)$. The orange and blue lines are for bands with $\mc{P}_1 = \pm 1$
respectively, both with negative energy. (c) The diagonal winding number 
defined in Eq.~\eqref{eq_mirror_winding2}. In (d) we show the parametric plot 
of $b(k_x)$ vs. $a(k_x)$ as described in Eq.~\eqref{abk}. The diagonal winding 
number is defined as the number of times this curve winds around $(0,0)$.
In this figure we have chosen $M = 0.1$, and we find $\ga = 1$. 
This is consistent with the phases described in (c).}
\end{center}
\end{figure}
\end{widetext}

We now consider the spectrum $E(\bk)$ obtained from Eq.~\eqref{eq:con_ham} 
for $\De_2=0$. Since there is a twofold degeneracy, for each value of $\bk$, 
there are only two distinct energy bands corresponding to positive and 
negative energies; the exact expressions for $E(\bk)$ will be presented below.
(We will present numerical results only for the negative-energy bands, 
but with $\mc{P}_1$ equal to both $+1$ and $-1$. The states in these bands 
can be found by projecting with the operators $(\mathbb{I}_4 \pm \mc{P}_1)/2$ 
respectively). The gap between the two bands is shown in Fig.~\ref{fig:E_gap}
as a surface plot. For a nonzero value of $\De_1$, the spectrum is found to 
be gapless for $M=0, \pm 2$. The region $|M| > 2$ is topologically trivial and
the Chern number is zero as shown in Fig.~\ref{fig:chern_eq}. (The Chern number
is computed using the method described in Ref.~\onlinecite{fukui}; the
Berry curvature is calculated as a function of $(k_x,k_y)$ and the Chern
number is then obtained by integrating over the Brillouin zone. 
We note that the bands with $\mc{P}_1 = \pm 1$ have opposite values 
of the Chern numbers).
The regions $-2 < M < 0$ and $0 < M < 2$, labeled as $A$ and $B$ in the figure,
are both topological. The Chern numbers in these two regions are nonzero and 
have opposite signs. By studying the Hamiltonian for an infinitely long ribbon 
(in Sec.~\ref{sec:st_ribbon}), we find there there are robust one-dimensional 
gapless edge states in the topological regime as shown in 
Fig.~\ref{fig:E_kx_noD2}.

Now switching on the $\De_2$ term, we find that the symmetries of 
the Hamiltonian get reduced as follows.

\begin{enumerate}

\item For $\De_2 \ne 0$, the time-reversal symmetry $\mc{T}$ as well 
as the rotational symmetry $\mc{C}_4$ are broken. However the symmetry given 
by the product of the two, i.e., $\mc{C}_4 T$ is preserved. This means that 
$(\mc{C}_4 \mc{T})H (\bk)(\mc{C}_4 \mc{T})^{-1} = H (\mc{C}_4 T \bk)$,
where $\mc{C}_4 T (k_x,k_y) = (-k_y,k_x)$.

\item The charge conjugation and one of the other symmetries act as before:
$\mc{C} H (\bk) \mc{C}^{-1} = -H^{*} (-\bk)$ and $\mc{S}_2 H (\bk) 
\mc{S}_2 =-H (\bk)$. The operator $\mc{S}_1$ does not describe a 
symmetry of the system.

\end{enumerate}

For $\De_2 \ne 0$, we cannot use $\mc{P}_1$ to distinguish between the 
two degenerate bands since $\mc{P}_1$ does not commute with the Hamiltonian.
It is therefore not possible to calculate the Chern number when $\De_2 \ne 0$.

We note that in the presence of a $\De_2$ term, the $\mc{C}_4 T$ symmetry gaps 
out the edge states but gapless corner states appear for certain values of $M$.
The system is then called a second-order TI.

The energy spectrum of the Hamiltonian in Eq.~\eqref{eq:con_ham} can be 
found as follows. Since the four matrices $\tau^z \otimes \si^0$, $\tau^x 
\otimes \si^x$, $\tau^x \otimes \si^y$ and $\tau^y \otimes \si^0$ anticommute 
with each other and the square of each of them is equal to $\mathbb{I}_4$, the
spectrum can be found by taking the square of $H(\bk)$. This gives
\bea E(\bk) &=& \pm ~[ (M ~+~ t_0 ~(\cos k_x ~+~ \cos k_y))^2 \non \\
&& ~~~~+~ \De_1^2 ~(\sin^2 k_x ~+~ \sin^2 k_y) \non \\
&& ~~~~+~ \De_2^2 ~(\cos k_x ~-~ \cos k_y)^2]^{1/2}. \label{spec1} \eea
(The fact that the Hamiltonian is a $4 \times 4$ traceless matrix and 
there are only two possible energy levels implies that each energy level 
must be twofold degenerate). Equation~\eqref{spec1} implies that the bulk 
gap can only vanish at one of the four 
momenta $\bk =(0,0), ~(0,\pi), ~(\pi,0)$ and $(\pi,\pi)$; further the energy 
at one of these momenta is zero only if (i) $M = \pm 2 t_0$ or (ii) $M=0$ and 
$\De_2 = 0$. These parameter values give the locations of topological phase
transitions as we will see below. At all other values of the parameters, the 
spectrum will be gapped at all momenta.

\onecolumngrid

Based on the existence of mirror symmetries, certain mirror-graded 
topological 
invariants were derived in Refs.~\onlinecite{schindler}, \onlinecite{imhof} 
and \onlinecite{rodriguez2}. While our isotropic model has a similar mirror 
symmetry, the anisotropic model discussed in Sec.~\ref{sec:st_asym} does not 
have that symmetry. Nevertheless, we will see that both models allow us to
define a topological invariant called `diagonal' winding number (which is
motivated by the mirror-graded topological invariants mentioned above); we use
the word `diagonal' since this winding number is defined on one of the 
diagonals in the Brillouin zone as we will see. In order to derive this 
winding number, we observe that the Hamiltonian in Eq.~\eqref{eq:con_ham} 
simplifies considerably along the diagonals $k_x = \pm k_y$. For instance, 
along $k_x = k_y$, the Hamiltonian takes the form
\bea H (k_x = k_y) &=& \big[ M ~+~ 2 t_0 ~\cos k_x \big] ~\tau^z \otimes \si^0
\non \\
&&+ ~\sqrt{2} \De_1 ~\sin k_x ~\tau^x \otimes \frac{\si^x ~+~ \si^y}{\sqrt 2},
\label{eq:con_ham2} \eea
which, interestingly, does not depend on $\De_2$. Next, we see that by
rotating by an appropriate angle using the matrix $\si^z$, we can transform
$(\si^x + \si^y)/\sqrt{2} \to \si^x$. We then obtain
\bea H (k_x=k_y) &=& \big[ M ~+~ 2 t_0 ~\cos k_x \big] ~\tau^z \otimes \si^0
\non \\
& &+ ~\sqrt{2} \De_1 ~\sin k_x ~\tau^x \otimes \si^x. \label{eq:con_ham3} \eea
This Hamiltonian only depends on two matrices, $\tau^z \otimes \si^0$
and $\tau^x \otimes \si^x$, which are Hermitian and square to $\mathbb{I}_4$.
If we take the coefficients of the two matrices,
\bea a(k_x) &=& M ~+~ 2 t_0 ~\cos k_x, \non \\
b(k_x) &=& \sqrt{2} \De_1 ~\sin k_x, \label{abk} \eea
to be the coordinates of a point in a two-dimensional plane, we obtain a
closed curve as $k_x$ goes from $-\pi$ to $\pi$. Assuming that the system is
gapped for all values of $(k_x,k_y)$ which includes the line $k_x = k_y$
as a special case, we see that $(a(k_x),b(k_x))$ will not be equal to
$(0,0)$ for any value of $k_x$. We can therefore define the winding number
$\ga$ of the closed curve around the origin; this will be a topological 
invariant since it will not change if any of the parameters are changed 
slightly (assuming that such changes do not make the curve pass through the
origin). Mathematically, we have
\bea \ga &=& \int_{-\pi}^\pi ~\frac{dk_x}{2\pi} ~\frac{d \phi_{k_x}}{dk_x},
\non \\
{\rm where}~~~~ \phi_{k_x} &=& \tan^{-1} \left(\frac{b(k_x)}{a(k_x)} \right).
\label{eq_mirror_winding2} \eea
We note that the existence of the winding number is crucially dependent
on the fact that on the diagonals $k_x = \pm k_y$, the Hamiltonian essentially
reduces to a sum of only two anticommuting matrices. If the Hamiltonian
contained a third anticommuting matrix such as $\tau^x \otimes \si^z$ or 
$\tau^y \otimes \si^0$, it would not be possible to define the diagonal 
winding number. We also observe that the diagonal winding number does not 
depend on the value of $\De_2$, unlike the Chern number which can only be 
calculated if $\De_2 = 0$.

A parametric plot of $b(k_x)$ versus $b(k_x)$ is shown in Fig.~\ref{fig:mw1_eq}
for the case when $M=0.1$, $t_0 = 1$, and $\De_1 = 1$. We can see easily that
the diagonal winding number is 1. This is consistent with the lattice 
calculation for this set of parameters where we find one localized 
state at every corner.

Figure~\ref{fig:mirror_eq} shows the diagonal winding number for the model 
described in Eq.~\eqref{eq:con_ham}. Comparing this with 
Fig.~\ref{fig:chern_eq}, we see that the regions of nonzero values of the 
Chern number and diagonal winding number coexist in this model when $\De_2 = 0$.

\subsection{Edge states for an infinitely long ribbon}
\label{sec:st_ribbon}

We will first study the effect of the $\De_2$ term on the edge states in this 
system. To this end, we consider a strip of the material which is infinitely 
long in the $x$ direction and has a finite width (with $N_y$ sites) in the 
$y$ direction. This means that $k_x$ is a good quantum number and for each 
$k_x$, we effectively have a chain with $N_y$ sites which is related to its 
neighboring chains by factors of $\exp(\pm i k_x)$. At each site there are 
four degrees of freedom corresponding to the two orbitals ($d$ and $f$ which 
we denote by $\tau^z = \pm 1$) and two spins ($\ua$ and $\da$ denoted
by $\si^z = \pm 1$). The creation operator for the $d$ orbital at 
the $n_y$-th site of the chain for spin $s$ is denoted by $d^\dag_{n_y,s}$ 
where $s = \ua, \da$. Similarly, for the $f$ orbital, the corresponding 
operator is $f^\dag_{n_y,s}$. Hence the spinor for each value of $k_x$ has 
$4N_y$ components and the Hamiltonian $H(k_x)$ is a $4N_y \times 4N_y$ matrix. 
By keeping the $k_x$ part of the Hamiltonian as it is, but 
discretizing in the $y$ direction, we obtain the following Hamiltonian,
\begin{subequations}
\bea H (k_x)&=& h^0_{k_x} \non \\
&& +\sum_{\substack{n_y=1\\s,\bs=\ua,\da}}^{N_y-1} \Bigg[\frac{t_0}{2} 
\Big(d^\dag_{n_y+1,s} d_{n_y,s} -f^\dag_{n_y+1,s} f_{n_y,s}\Big)\non \\
&& +\frac{\De_1}{2} \eta_{s,\bs} \Big ( d^\dag_{n_y+1,s} f_{n_y,\bs} + 
f^\dag_{n_y+1,s} d_{n_y,\bs} \Big) \non \\
&& -\frac{i \De_2}{2} \Big(f^\dag_{n_y+1,s} d_{n_y,s} + f^\dag_{n_y,s} 
d_{n_y+1,s} \Big) + H.c.\Bigg], \non \\ \label{eq:H_ribbon}
\eea
where $\eta_{s,\bs} = \pm 1$ for $s = \ua(\da)$, and $h^0_{k_x}$ is a 
$4N_y \times 4N_y$ matrix given by
\bea
h^0_{k_x} &=& \mathbb{I}_{N_y} \otimes \Big((M ~+~ t_0\cos k_x) ~\tau^z 
\otimes \si^0 \non \\
&& ~~~~~~~~~+~ \De_1 \sin k_x ~\tau^x \otimes \si^x \non \\
&& ~~~~~~~~~+~ \De_2 \cos k_x ~\tau^y \otimes \si^0 \Big), \label{eq:h_kx} 
\eea \label{eq:Ham_ribbon}
\end{subequations}
where $\mathbb{I}_{N_y}$ is the $N_y \times N_y$ identity matrix. \\

\begin{figure}[htb]
\begin{center}
\sg[~$\De_2 =0$]
{\ig[height=4cm]{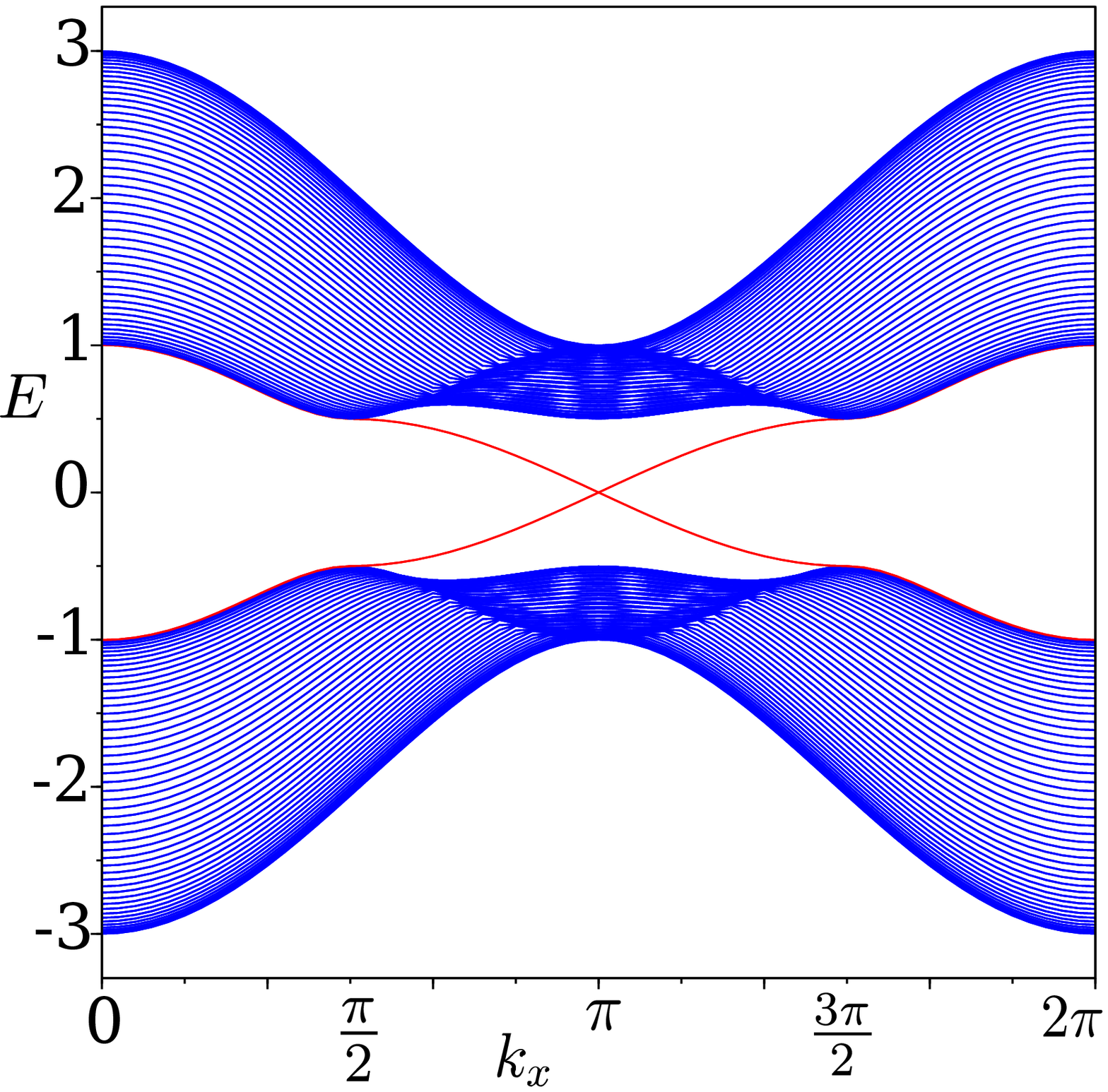}\label{fig:E_kx_noD2}}
\hspace{0.1cm}
\sg[~$\De_2=0.1$]{\ig[height=4cm]{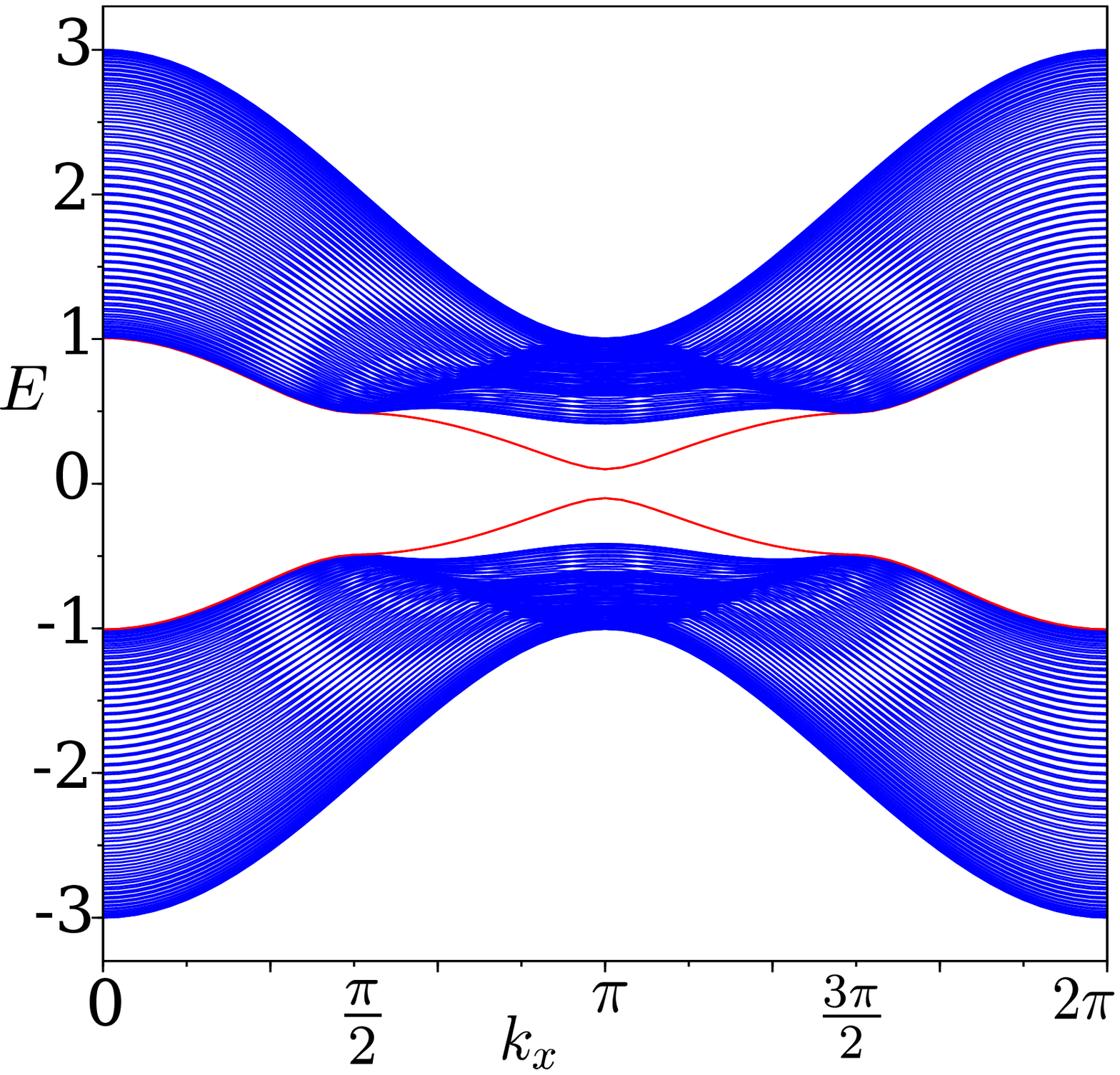}\label{fig:E_kx_D2}}
\end{center}
\caption[Ribbon Geometry]{Edge and bulk dispersions for (a) $\De_2 =0$ 
and (b) $\De_2 = 0.1$. The blue regions in both the figures correspond to the 
bulk states. The solid red curves lying in the gap are one-dimensional edge 
states whose wave functions decay exponentially into the two-dimensional bulk 
and are doubly degenerate. In (a), these edge states go from one band to the
other making the system topological, and their number is related to the Chern 
number. In (b) the $\De_2$ term gaps out these edge states which now lie within
the same band and are therefore nontopological.} \label{fig:E_kx} \end{figure}

In our numerical calculations, we will set $t_0 = 1$ and express all the
other parameters in units of $t_0$. Diagonalizing Eq.~\eqref{eq:H_ribbon}, we 
find $4N_y$ energy eigenvalues and eigenvectors for each momentum $k_x$, such 
that each eigenvector is a $(4N_y)$-component spinor. The energy levels as a 
function of momentum $k_x$ are shown in Fig.~\ref{fig:E_kx} for both 
$\De_2 =0 $ and $\De_2 \neq 0$. We have taken $M=1$ and $\De_1 = 0.5$.

The bulk states form the two continuous energy bands shown in blue. These two 
bands are separated by an energy gap whose value depends on both $M$ and 
$\De_1$ consistent with Fig.~\ref{fig:E_gap}. Let us first consider 
Fig.~\ref{fig:E_kx_noD2} where we have taken $\De_2=0$. Here the bulk gap hosts
gapless edge states (red solid lines) which go from one band to the other and 
are related to the Chern number of the infinite system by the bulk-boundary 
correspondence; we thus have a two-dimensional TI. On 
switching on a $\De_2$ term, these edge states become gapped as shown in 
Fig.~\ref{fig:E_kx_D2}. Since each of these edge states lie within the same 
band, i.e., they do not connect the two bands, they are topologically trivial.
This is consistent with the fact the Chern numbers are zero for the infinite 
system with $\De_2 \neq 0$. A useful topological invariant for $\De_2 \neq 0$ 
is the diagonal winding number $\ga$ which is explained in 
Sec.~\ref{sec:st_bulk}. However, the bulk-boundary correspondence for this 
case has to be analyzed using a system which is finite in both the directions, 
such as a finite-sized square lattice, as discussed in the next section. 

\subsection{Corner states for a finite-sized square lattice}
\label{sec:st_corner}

In order to find the corner states of the system, we consider a finite-sized
square lattice lying in the $x-y$ plane with $N_x$ and $N_y$ sites along the 
$x$ and $y$ directions respectively. The total number of lattice points 
is $N = N_x\times N_y$. Since there are two orbital ($d$ and $f$) and two 
spin ($\ua$ and $\da$) degrees of freedom at each site, we arrive at a $4N 
\times 4N$ Hamiltonian, 
\bea H &=& \sum_{\substack{\la n,n'\ra \\s=\ua,\da}}
\Bigg[ \frac{M}{2} \Big(d_{n,s}^\dag d_{n,s}^{}-f_{n,s}^\dag f_{n,s}^{}
\Big)\non\\
&& ~~~+ ~\frac{t_0}{2} \Big( d_{n,s}^\dag d_{n',s}^{} - f_{n,s}^\dag 
f_{n',s}^{} \Big) \non \\
&& ~~~+~\frac{\De_1}{2} \eta_{s,\bs}~\ep_{n,n'} \Big(d_{n,s}^{\dag} 
f_{n',\bs}^{} + f_{n,s}^{\dag}d_{n',\bs}^{}\Big) \non \\
&& ~~~+ ~\frac{i \De_2}{2} \xi_{n,n'} \Big(d_{n,s}^{\dag} f_{n',s}^{} - 
f_{n,s}^{\dag}d_{n',s}^{}\Big) + H.c. \Bigg], \non\\ 
\label{eq:Ham_sq} \eea
where ${\la n,n'\ra}$ denotes nearest neighbors, $\ep_{n,n'} = i$ or 1 
and $\xi_{n,n'} = \pm 1 $ for nearest neighbors along the $x$ and 
$y$ directions respectively. 

The different terms in Eq.~\eqref{eq:Ham_sq} can be understood as follows. $M$ 
acts as a staggered chemical potential for the two orbitals. $t_0$ is the 
amplitude for nearest-neighbor hopping that keeps both the spin and orbital 
the same but differs in sign for the two orbitals. $\De_1$ flips both the spin
and orbital degrees of freedom and also depends on the direction of hopping 
via $\ep_{n,n'}$. $\De_2$ describes a hopping that flips the orbital 
but keeps the spin the same, and this term depends on the direction of hopping 
through $\xi_{n,n'}$. Diagonalizing this Hamiltonian gives the $4N$ energy 
eigenvalues and eigenvectors, all of which turn out to be doubly degenerate.

\begin{widetext}
\begin{figure}[h]
\begin{center}
\sg[~Energy levels]{\ig[height=4.6cm]{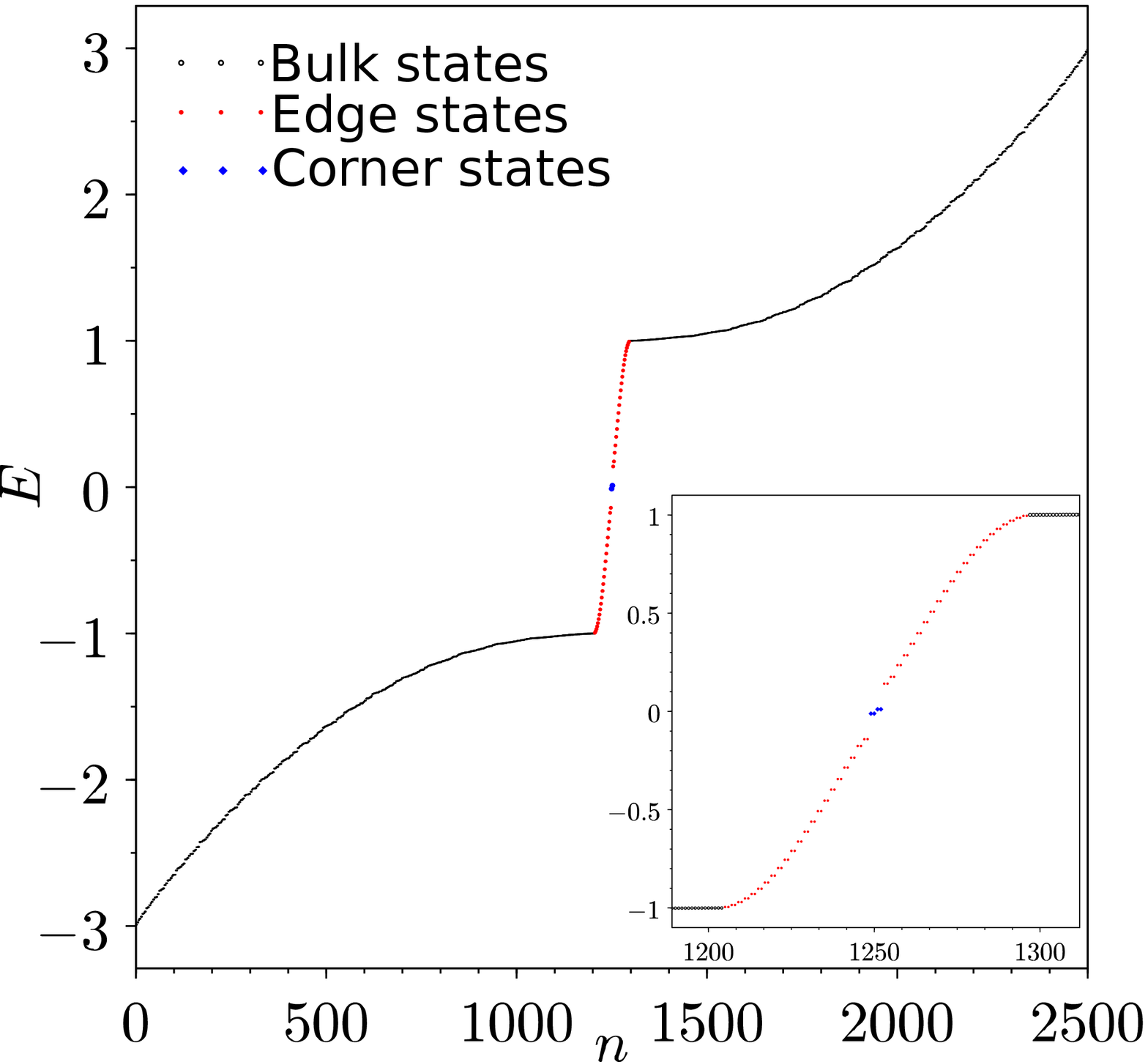}
\label{fig:Elevels}}
\hspace{0.5cm}
\sg[~Corner state]{\ig[height=4.6cm]{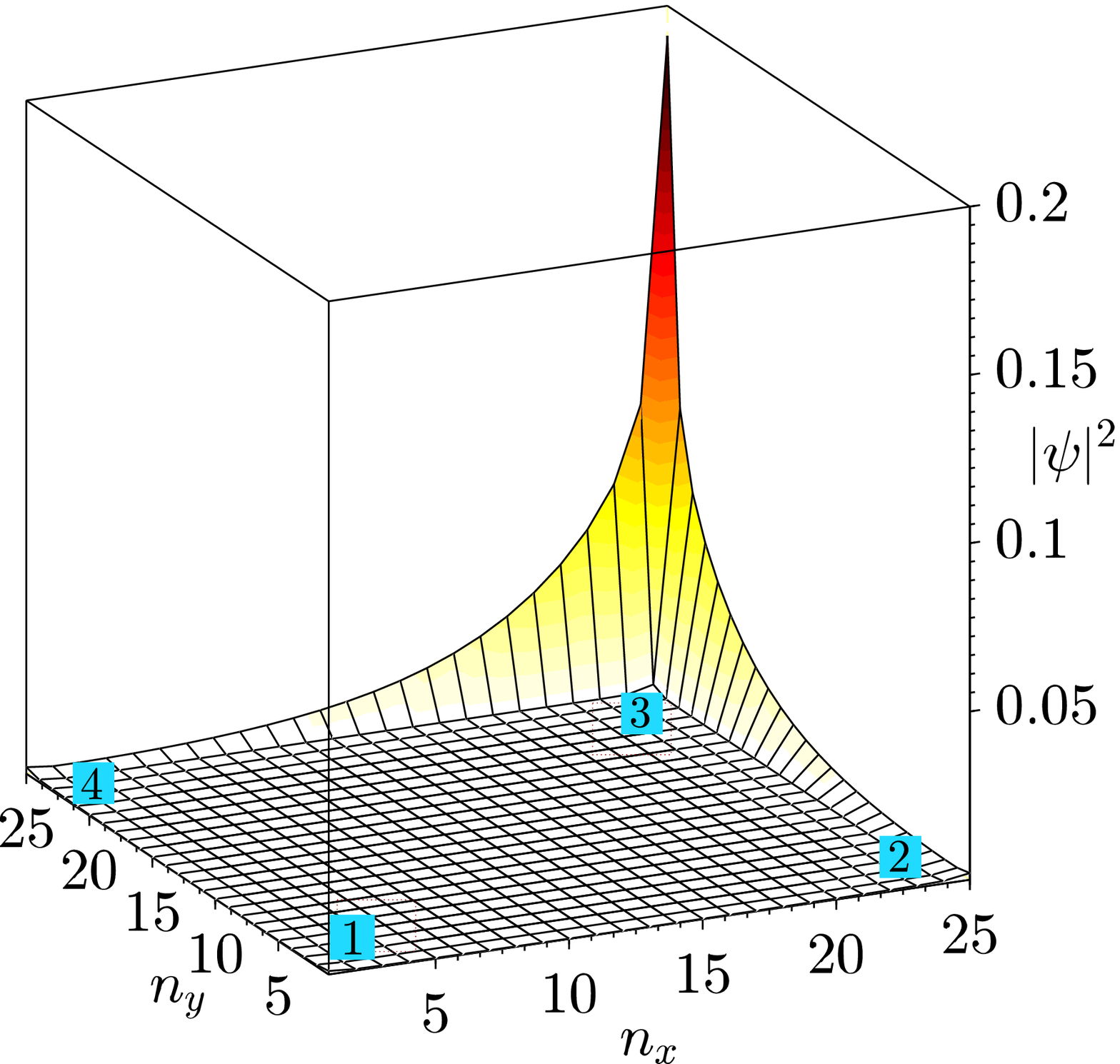}\label{fig:st_corner}}
\hspace{0.4cm}
\sg[~Edge state]{\ig[height=4.6cm]{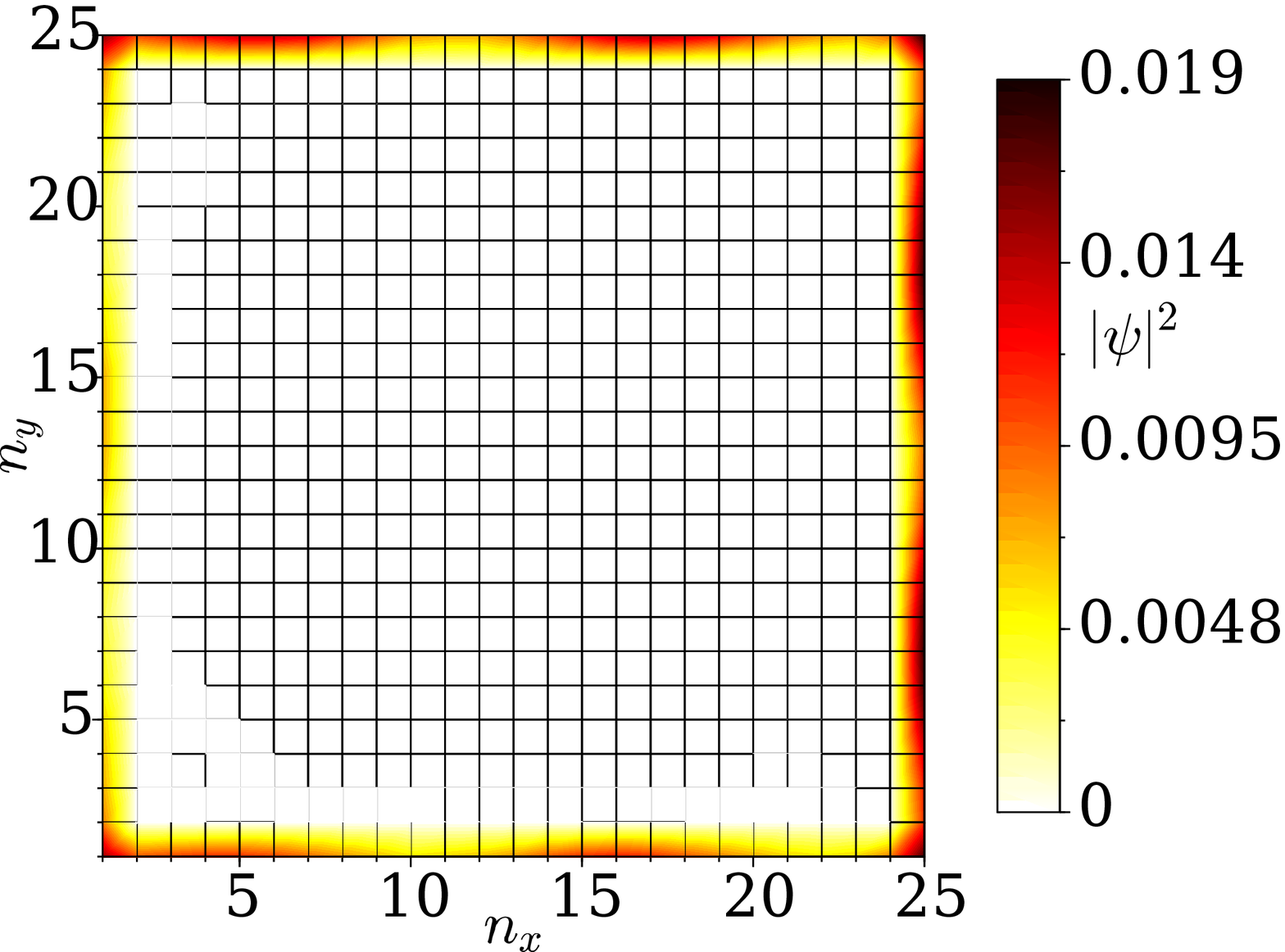}\label{fig:st_edge}}
\end{center}
\caption[Bulk, edge and corner state energies]{(a) Energies of bulk states 
(black), edge states (red) and corner states (blue) for a lattice with
$25 \times 25$ sites. The inset in (a) shows the magnified version. There are 
four corner states close to zero energy which become degenerate in the 
thermodynamic limit. By appropriately superposing these degenerate 
zero-energy states, we obtain an eigenfunction shown in (b) which has a high 
probability at exactly one corner and decays exponentially into the bulk. 
This decay is very sharp along the diagonal into the bulk and is more gradual 
along the edges. (c) An edge state which lives on the boundary of the system. 
The energy of the state is $0.5$ in units of $t_0$.}
\label{fig:static} \end{figure}
\end{widetext}

For our numerical calculations, we have considered a square lattice with $25$ 
sites in each direction, i.e., $N_x = N_y = 25$. The parameter $t_0$ is set 
to unity and all other parameters and the energy are expressed in units of 
$t_0$. The results for $M=1$, $\De_1 = 1$ and $\De_2 = 0.1$ are shown in 
Fig.~\ref{fig:static}. The plot of energy eigenvalues versus the eigenvalue 
index is shown 
in Fig.~\ref{fig:Elevels}. The edge states (red) and corner states (blue) are
clearly separated in energy from the bulk states (black). As is clear from the 
inset in this figure, there are four states which are very close to zero 
energy. They become degenerate in the thermodynamic limit; these fourfold 
degenerate states are found to exist only at the corners of the square. 
One such state which lives at the corner labeled $3$ is shown in 
Fig.~\ref{fig:st_corner}. Similar states at zero energy exist at each of the 
four corners. We find that the decay length of these states is much larger 
along the edges than along the diagonal direction into the bulk. This can 
be understood as follows. Typically, the decay length of a boundary state is 
inversely proportional to the gap of the corresponding bulk states. In our 
system, the bulk gap is much larger than the gap of the edge states (the 
latter is proportional to $\De_2$); we can clearly see this difference in 
Fig.~\ref{fig:Elevels}. Hence the decay length into the bulk is 
much smaller than the decay length along the edges.

We have found that the four-component spinors corresponding to these corner 
states have an interesting structure. We recall that the Hamiltonian in 
Eq.~\eqref{eq:con_ham} changes sign under a transformation by $\mc{S}_2 = 
\tau^x \si^z$. We therefore expect the space of zero-energy states to remain 
invariant under the action of $\mc{S}_2$; in particular, the corner states 
should be eigenstates of $\mc{S}_2$. We find that the states localized at 
corners labeled $1$ and $3$ have eigenvalues of $\mc{S}_2$ equal to $+1$ 
while those at corners $2$ and $4$ have eigenvalues $-1$. We can 
understand the relation between the eigenvalues of $\mc{S}_2$ for the states 
at the different corners as follows. First, Eq.~\eqref{eq:con_ham} implies 
that a rotation by $\pi$ about
the $z$ axis (which transforms $x \to -x$ and $y \to - y$) can be performed
by the operator $\si^z$. This transformation interchanges the corners (1,3)
and corners (2,4). Since $\si^z$ commutes with $\mc{S}_2$, this implies
that the eigenvalues of $\mc{S}_2$ for the states at corners 1 and 3 must
be identical, and similarly for the eigenvalues for the states at corners 2
and 4. Second, we see from Eq.~\eqref{eq:con_ham} that the reflection $x \to 
-x$ and $y \to y$ can be performed by the operator $\si^y$. This
interchanges the corners (1,2) and corners (3,4). Since $\si^y$ anticommutes
with $\mc{S}_2$, this implies that the eigenvalues of $\mc{S}_2$ for the
states at corners 1 and 2 must have opposite signs, and similarly for the
eigenvalues for the states at corners 3 and 4.

The edge states, whose energies are shown in red in Fig.~\ref{fig:Elevels}, are
found to be localized along all the edges of the system. However, these are not
protected topologically as explained in the analysis of the ribbon geometry in 
Sec.~\ref{sec:st_ribbon} (they do not go from one bulk band to the other). 

\onecolumngrid

To recapitulate, in this section we have analyzed a topologically nontrivial 
two-dimensional system which, in certain parameter regimes is found to have 
robust zero-energy ``corner states". These can be connected to the bulk 
system using a topological invariant called the diagonal winding number.

It is useful to understand why the $\De_2$ term in Eq.~\eqref{eq:con_ham}
gaps out the edge states and produces zero-energy corner states. Basically,
the $\De_2$ term provides a mass term for the edge states and therefore
gaps them out. However, the mass term has opposite signs for edges running
along the $x$ and $y$ directions, due to the $\cos k_x - \cos k_y$ structure.
Since the mass terms changes sign on going around a corner from an edge along 
$x$ to an edge along $y$, a zero-energy state appears at each 
corner~\cite{schindler}. This mechanism for the appearance of zero-energy 
corner states is different from the one discussed in 
Ref.~\onlinecite{matsugatani}. That work considers a model of a 
three-dimensional TI; the addition of a weak magnetic field gaps out the 
surface states but produces gapless states which are localized along the 
vertical hinges. Then the coupling in the $z$ direction is smoothly turned 
off; this produces a set of decoupled two-dimensional TIs each of which has 
zero-energy corner states which are the remnants of the hinge states.
We also note that our model lies in the symmetry class BDI while the one 
discussed in Ref.~\onlinecite{matsugatani} is in the class AIII.

\section{Periodically driven isotropic model}
\label{sec:driven}

Having understood the properties of the time-independent model of a 
second-order TI in various parameter regimes, we now proceed to study what 
happens when the system is driven periodically in time.
In particular, we will study the effect of varying the parameter $M$ in the 
Hamiltonian. We will study the topological properties of this driven system 
both in momentum space (i.e., for a system with periodic boundary conditions) 
as well as for a finite-sized square lattice using a time-evolution operator,
and we will again use the Chern number and diagonal winding number to
characterize the system~\cite{bomantara,huang,rodriguez2,peng2}.


First, let us consider the bulk system, i.e., we work with the momentum 
space Hamiltonian $H$ given in Eq.~\eqref{eq:con_ham} with the parameter $M$ 
varying in time as,
\beq M(t) = \begin{cases} ~~~ M_1 ~&~ \text{if} ~~~ 0 < t < T/4 \\ 
~~~M_2 ~&~ \text{if}~~~ T/4 < t < 3T/4 \\
~~~ M_1 ~&~ \text{if}~~~ 3T/4 < t < T \\
\end{cases}
\label{eq:Mt} \eeq
within a single time period $[0,T]$; we then continue this periodically
by taking $M(t+T)=M(t)$ for all $t$. 
We are interested in studying the system stroboscopically, i.e., at times 
$t = \mc{N} T$ where $\mc{N}$ runs over all integers. We will calculate
the quasienergy spectrum by numerically diagonalizing the Floquet operator 
$U_F$ (defined in Eq.~\eqref{eq:U_lat} below) for each momentum $\bk$. We 
find that each of the two quasienergy bands is twofold degenerate.

The time-evolution operator is given by 
\bea U_F &=& \mathscr{T}_t ~e^{-i\int_{0}^{T}H (t')dt'} \non \\
&=& e^{-iH_1 T/4} ~e^{-iH_2 T/2} ~e^{-iH_1 T/4}, \label{eq:U_lat} \eea
where $T$ is the time period of the driving and $H_1$ and $H_2$ are the 
Hamiltonians with $M = M_1$ and $M_2$ respectively.
(The symbol $\mathscr{T}_t$ denotes the time-ordered product). Since we will
study both momentum and real space systems, the Hamiltonian takes two 
different forms for these two cases, as will be explained below. Further,
since $U_F$ is a unitary operator, its eigenvalues are complex numbers with 
unit magnitude, i.e.,
\beq U_F |\psi_{j}\ra=e^{-i\ep_{j}T}|\psi_{j}\ra, \label{eq:U_floq} \eeq
where $\ep_{j}$'s are the quasienergies and are defined modulo $2\pi /T$. We 
define the first Floquet Zone such that $\ep_j \in [-\pi/T,
\pi/T]$ and $|\psi_{j}\ra$'s are the corresponding Floquet eigenstates.

The driving protocol described in Eq.~\eqref{eq:Mt} has been chosen
to satisfy a particular symmetry of the Floquet operator $U_F$. We saw
earlier that $\mc{S}_2 = \tau^x \si^z$ satisfies $\mc{S}_2 H \mc{S}_2^{-1}
= - H$; this is true in both momentum space and real space. 
Equation~\eqref{eq:U_lat} then implies that
\beq \mc{S}_2 ~U_F ~\mc{S}_2^{-1} ~=~ U_F^{-1}. \eeq
This implies that if $|\psi_{j}\ra$ is an eigenstate of $U_F$ with eigenvalue
$e^{-i\ep_{j}T}$, then $\mc{S}_2 |\psi_{j}\ra$ is an eigenstate of $U_F$ 
with eigenvalue $e^{+i\ep_{j}T}$. Thus the eigenvalues of $U_F$ must appear
in complex conjugate pairs. Next, if there is an eigenstate $|\psi_{j}\ra$
with eigenvalue $e^{-i\ep_{j}T} = \pm 1$, $\mc{S}_2 |\psi_{j}\ra$
must be the same as $|\psi_{j}\ra$. Hence $|\psi_{j}\ra$ must be an eigenstate
of $\mc{S}_2$, and the eigenvalue must be $\pm 1$ since $(\mc{S}_2)^2$ is the
identity operator. We will see below that corner states always appear with 
$e^{-i\ep_{j}T} = \pm 1$; hence they must also be eigenstates of $\mc{S}_2$ 
with eigenvalue $\pm 1$. Further, the arguments given in
Sec.~\ref{sec:st_corner} regarding the relative eigenvalues of
$\mc{S}_2$ at the four corners are valid for the periodically driven
system as well. Namely, a rotation by $\pi$ about the $z$ axis is
performed by $\si^z$ while a reflection of the $x$ coordinate is performed
by $\si^y$, and these commute and anticommute respectively with $\mc{S}_2$.
Combining these together, we see that if corner 1 has $(n_+,n_-)$ states
with eigenvalues of $\mc{S}_2$ equal to $(+1,-1)$ respectively, corner 3 will
have the same number of states with those eigenvalues, while corners 2 and 4
will have the number of eigenvalues of the two kinds interchanged to
$(n_-,n_+)$.


\vspace*{.2cm}

\begin{figure}[htb]
\begin{center}
\sg[~Chern number]{\ig[height=2.cm]{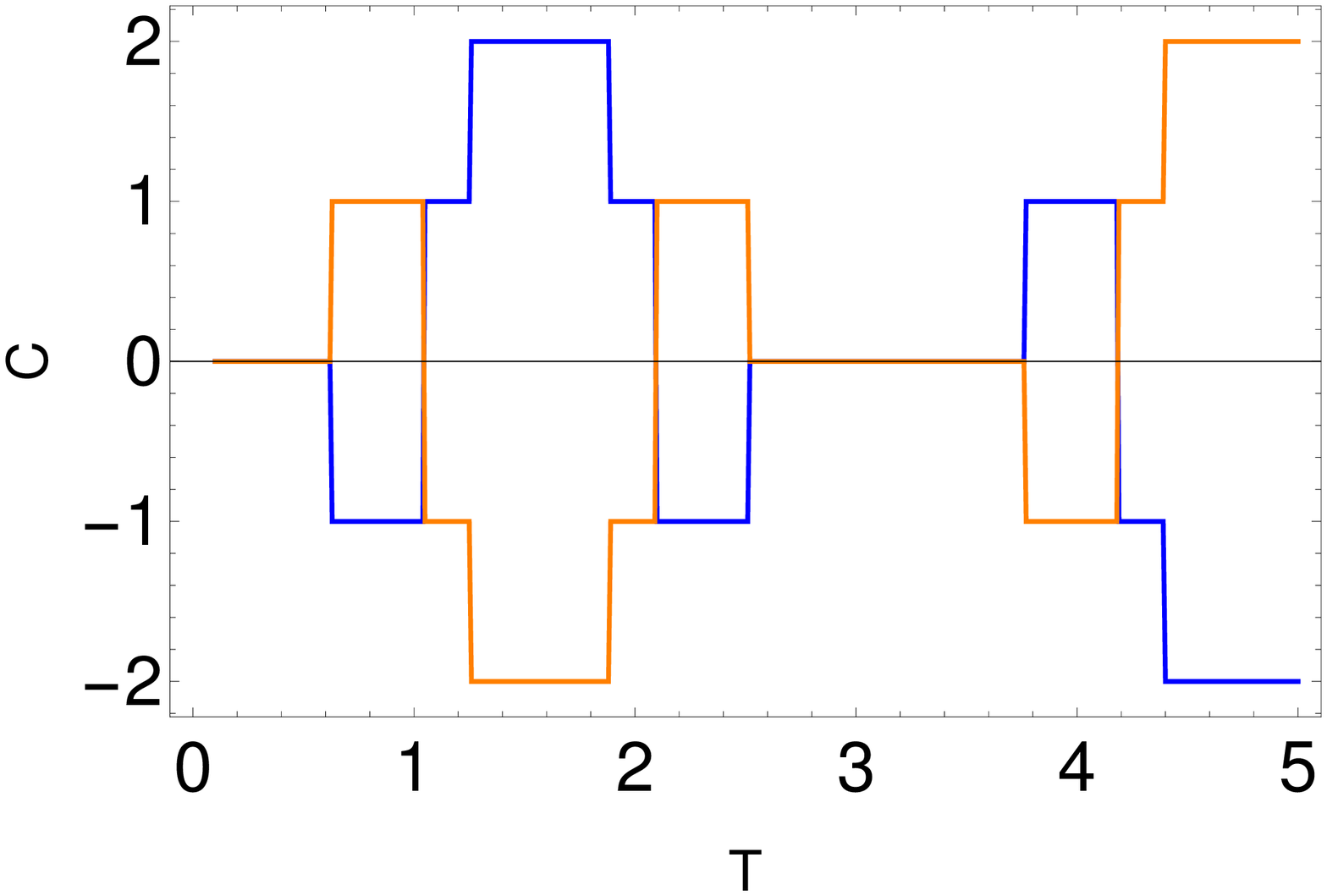} \label{fig:chern_dyn}}
\sg[~Winding number]{\ig[height=2.cm]{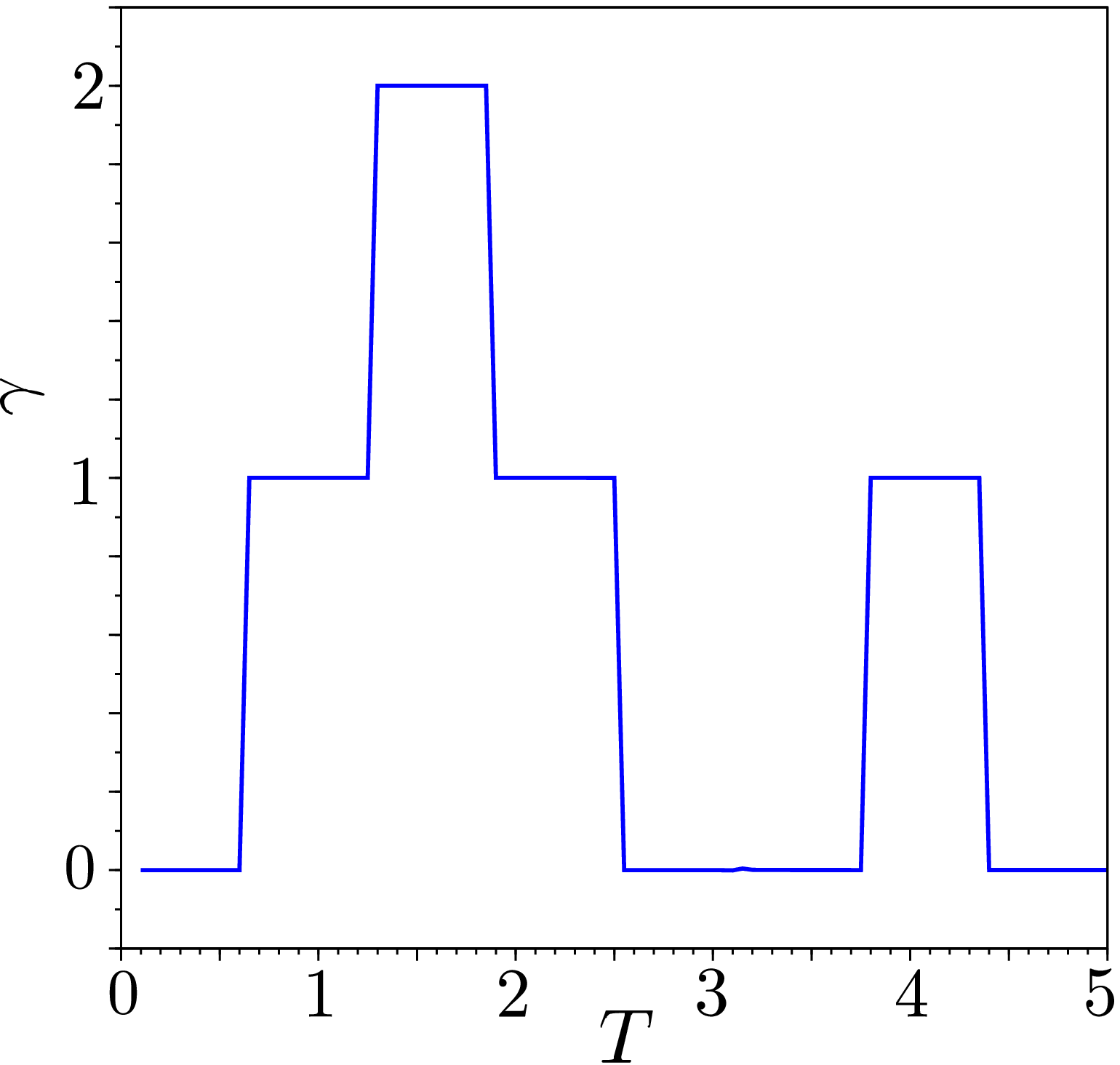} 
\label{fig:mirror_dyn}}
\sg[~$\ga = 2$]{\ig[height=2.cm]{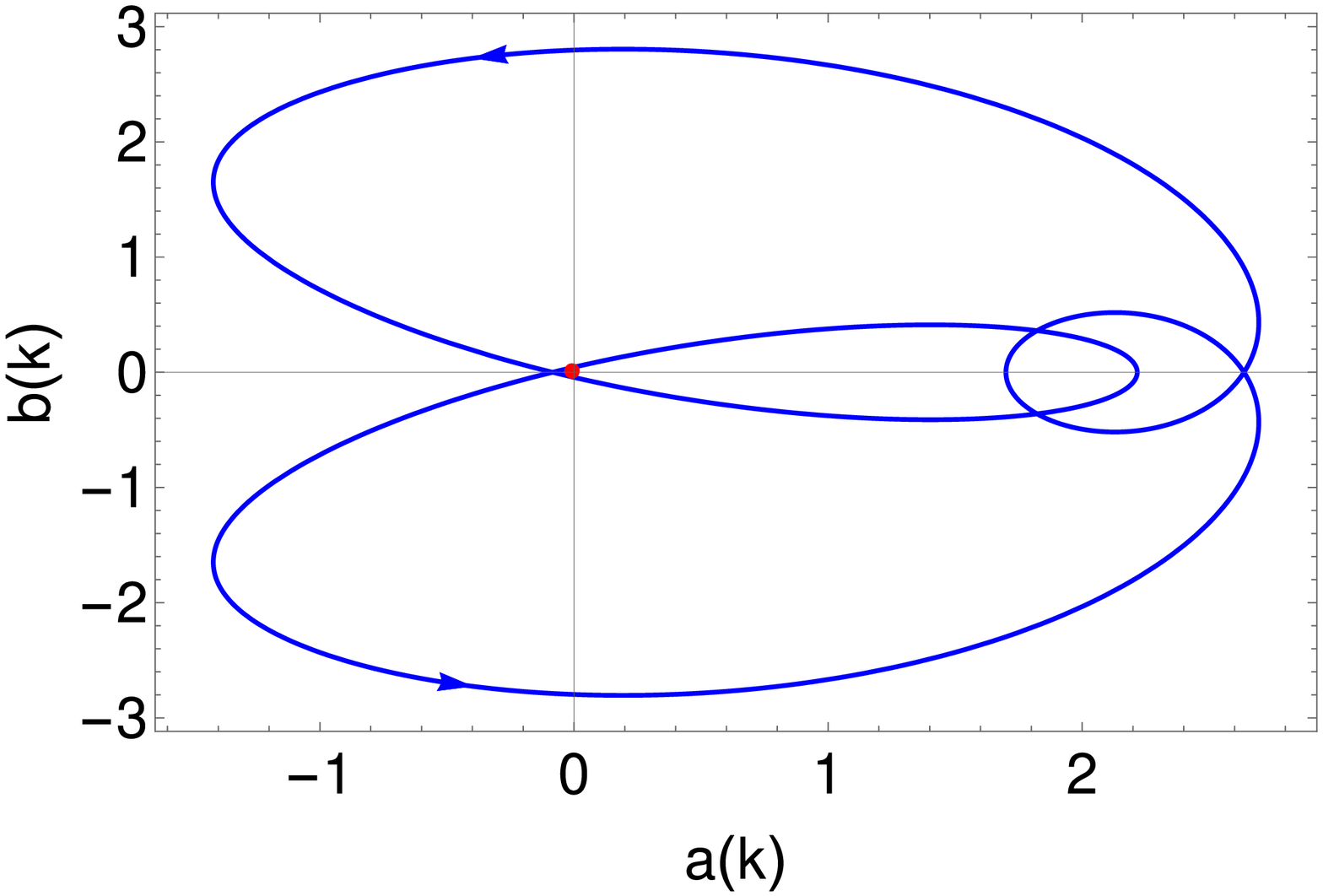}}
\end{center}
\caption{Topological invariants calculated from the Floquet 
operator defined in Eq.~\eqref{eq:U_lat} in momentum space for $\De_1 = 
1, ~M_1 = 2.5$ and $M_2 = 3.5$. Both the invariants 
are shown as functions of the time period $T$ defined in Eq.~\eqref{eq:Mt}. 
In (a) we show the dependence of the Chern number calculated from the
Floquet eigenstates for $\De_2 = 0$. Topological transitions from a topological
insulator to a nontopological insulator can be seen at certain values of 
$T$. In (b), we take $\De_2 = 0.1$ and plot the diagonal winding number $\ga$. 
The regions with nonzero values of $\ga$ are the Floquet second-order TI 
phases. In (c) we show a parametric plot of $b(k_x)$ vs. $a(k_x)$ defined in 
\eqref{eq:abk_fl} at $T=1.7$. As this curve goes around the origin twice in the 
anti-clockwise direction, we conclude that $\ga=+2$.} \label{fig:inv_dyn} 
\end{figure}


Having computed the Floquet operator $U_F (\bk)$, we investigate if there 
is any correspondence between the bulk states and the corner states that 
appear due to the driving. To choose an appropriate time period for studying 
the Floquet problem, we have to compare the time period $T$ with an intrinsic 
timescale of the system, which, in this case can be taken to be $1/t_0$. For 
fast driving, i.e., when the time period $T \ll 1/t_0$, the system does not 
have time to respond to the changing Hamiltonian and the properties are not 
very different from the static system. On the other hand, the system has 
interesting behavior for intermediate driving frequency, i.e., when $T$ is 
comparable to $1/t_0$. 

To study the intermediate frequency regime, we fix $M_1=2.5$ and $M_2 = 3.5$
and numerically find the eigenvalues and eigenvectors of $U_F (\bk)$ for 
different values of $T \approx 1/t_0$. 
We then use these eigenvectors to compute the topological invariants. 
The results for this are shown in Fig.~\ref{fig:inv_dyn}. The two topological 
invariants i.e., the Chern number and diagonal winding number are shown for the 
cases $\De_2 = 0$ and $\De_2 = 0.1$ in Figs.~\ref{fig:chern_dyn} and 
\ref{fig:mirror_dyn} respectively.

In Fig.~\ref{fig:chern_dyn} we have shown the Chern number for the case 
$\De_2=0$. The locations and magnitudes of the jumps in the Chern number can 
be understood by studying the time-evolution operator at some special points 
in the Brillouin zone. The Chern number changes abruptly when $U_F (\bk)$ 
becomes equal to $\mathbb{I}_4$ at one of the four time-reversal invariant 
momenta $(0,0)$, $(0,\pi)$, $(\pi,0)$ and $(\pi,\pi)$. 
The magnitude and sign of the jump is determined by the momentum 
point where the time-evolution operator becomes $\mathbb{I}_4$.
A detailed explanation of this is given in Appendix A.

We can also define the diagonal winding number for the periodically
driven system. Along one of the diagonals, say, $k_x=k_y$, we saw
in Sec.~\ref{sec:st_bulk} that, after an appropriate transformation, the
Hamiltonian becomes a linear combination of two matrices, $\tau^z \otimes
\si^0$ and $\tau^x \otimes \si^x$, at any time $t$. The commutator of these 
two matrices gives a third matrix, $\tau^y \otimes \si^x$, such that the three
matrices form a closed Lie algebra. The Floquet operator
$U_F (k_x=k_y)$ defined in Eq.~\eqref{eq:U_lat} must therefore be an
exponential of a linear combination of these three matrices. Next, we observe
that $\tau^z \otimes \si^0$ and $\tau^x \otimes \si^x$ are symmetric matrices
while $\tau^y \otimes \si^x$ is antisymmetric. Since $H_1$ and $H_2$ are
symmetric, $U_F (k_x=k_y)$ given by Eq.~\eqref{eq:U_lat} must also be a
symmetric matrix. Hence it must be the exponential of a linear combination
of only $\tau^z \otimes \si^0$ and $\tau^x \otimes \si^x$, and not
$\tau^y \otimes \si^x$. Thus we can write
\beq U_F (k_x=k_y) ~=~ \exp [i (a(k_x) \tau^z \otimes \si^0 ~+~ b(k_x)
\tau^x \otimes \si^x)], \label{eq:abk_fl}\eeq
where $(a(k_x),b(k_x))$ are fixed uniquely by demanding that $0 < \sqrt{
(a(k_x))^2 + (b(k_x))^2} < \pi$; we can demand this if $U_F (k_x=k_y)$ is 
not equal to $\pm \mathbb{I}_4$ for any value of $k_x$~\cite{top15}.
We now take $(a(k_x),b(k_x))$ to be the coordinates of a point in a
two-dimensional place, thus defining a closed curve as $k_x$ goes from $-\pi$
to $\pi$. We then define the winding number of the closed curve around the
origin as in Eq.~\eqref{eq_mirror_winding2}.


For the finite-sized system on a $25 \times 25$ lattice, we find that corner 
states only appear when the diagonal winding number of the bulk system is 
nonzero, implying that there is a nontrivial second-order bulk boundary 
correspondence. The time-evolution operator $U$ for the lattice model is given 
by the time-ordered product given in Eq.~\eqref{eq:U_lat}
with the Hamiltonian $H$ of the form given in Eq.~\eqref{eq:Ham_sq} with the 
parameter $M$ given by Eq.~\eqref{eq:Mt} for the two halves of the cycle, i.e.,
$H_1 = H|_{M = M_1}$ and $H_2=H|_{M = M_2}$. For a finite-sized square lattice, 
$U_F$ is a $4N \times 4N$ matrix where $N = N_x \times N_y$. The factor of $4$ 
is from the two spin and two orbital degrees of freedom at each lattice site. 
Diagonalizing this $U_F$ gives $4N$ quasienergy eigenvalues denoted by $\ep_j$
in Eq.~\eqref{eq:U_floq} and each eigenvector $|\psi_j\ra$ is a $4N$-component 
spinor. 

\begin{figure}[htb]
\begin{center}
\sg[~Quasienergy]{\ig[height=3.7cm]{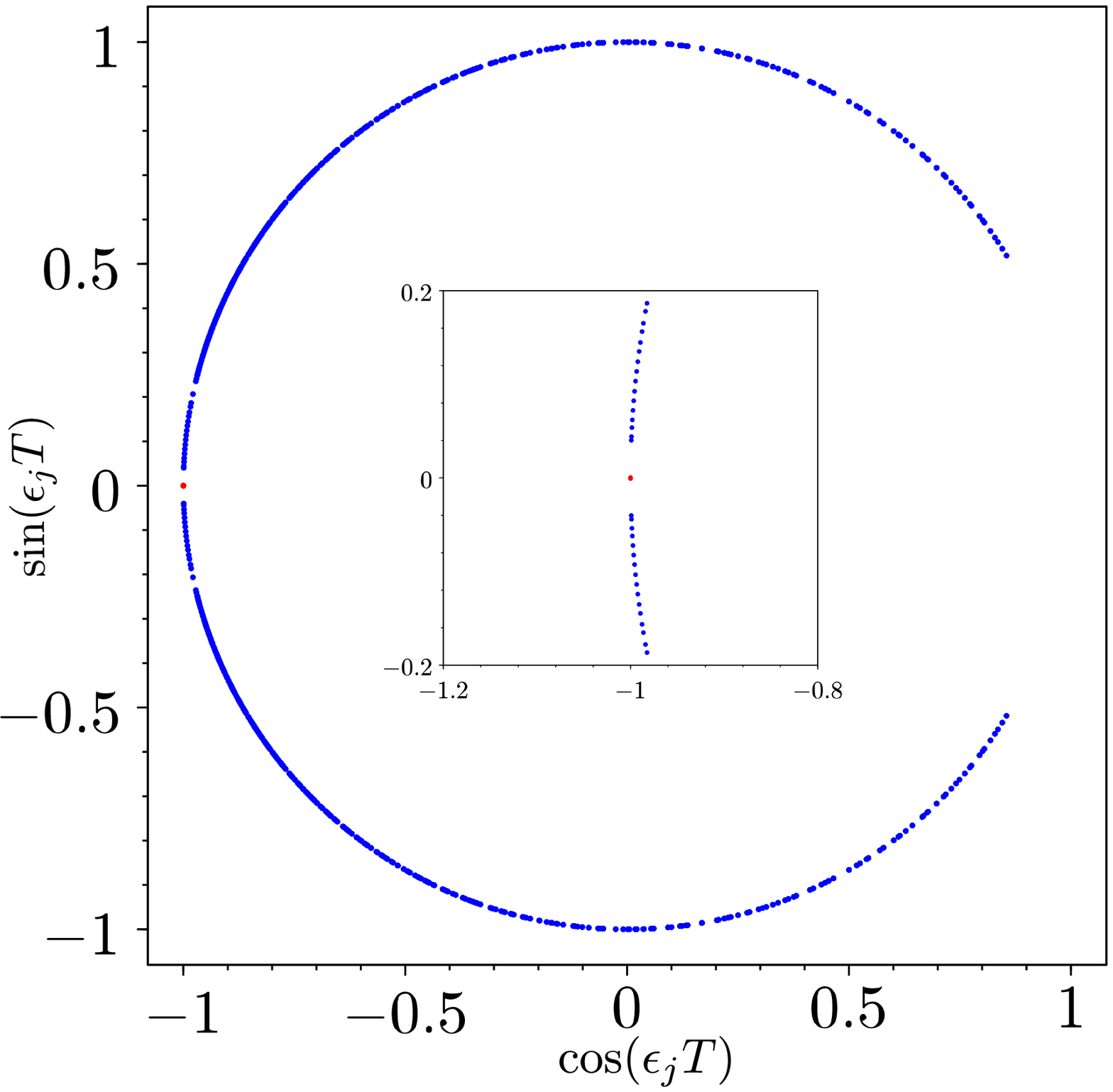}\label{fig:qE}}
\hspace{0.3cm}
\sg[~Corner state]{\ig[height=3.7cm]{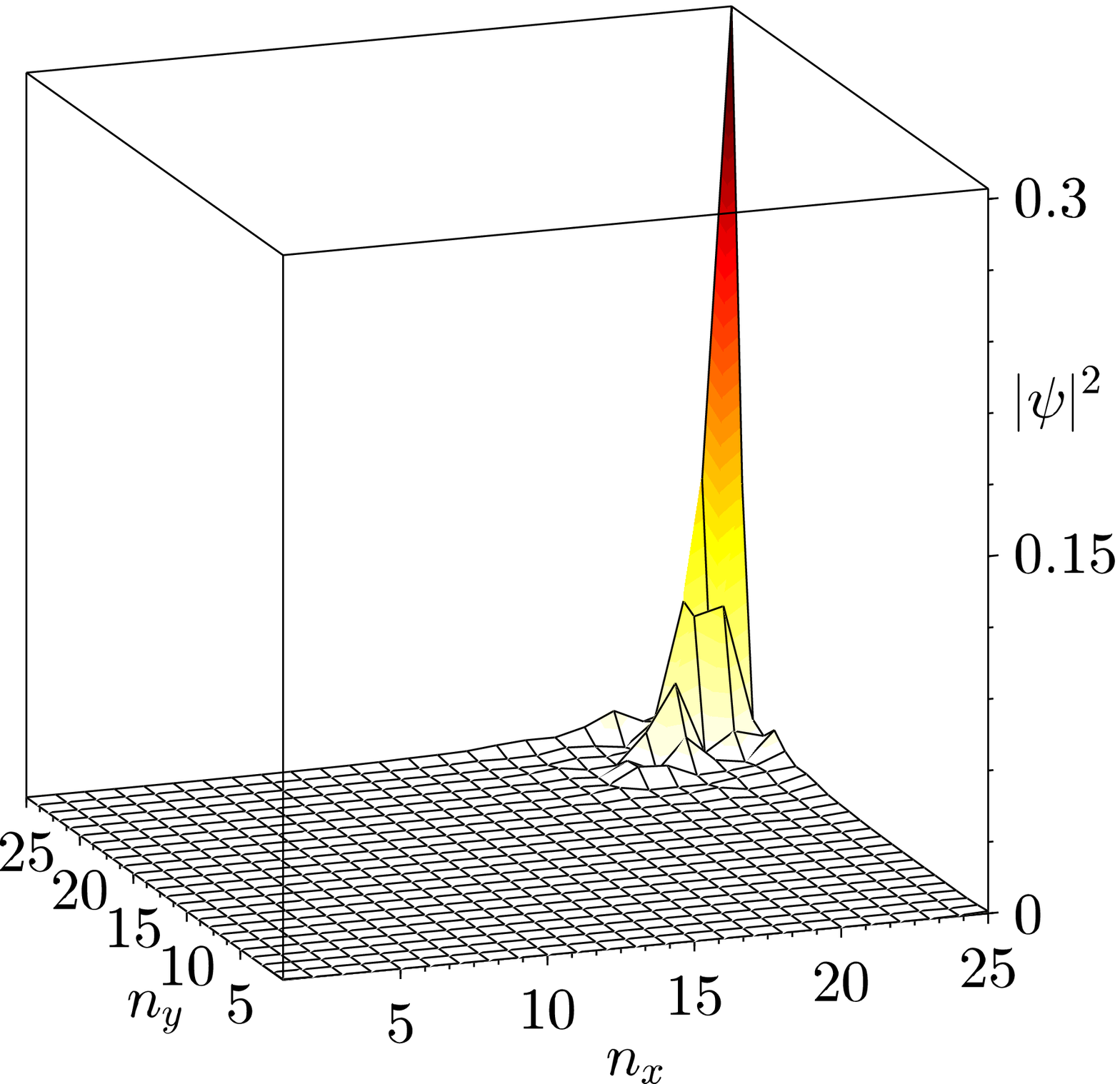}
\label{fig:corner_dyn}}
\end{center}
\caption[]{(a) Quasienergy spectrum and (b) corner state of the system driven 
as given by Eq.~\eqref{eq:Mt}. The red dots close to $(-1,0)$ in (a) are the 
quasienergies corresponding to the corner states. The corner state shown in (b) 
is obtained by an appropriate superposition of these four degenerate states. 
Here the results are shown for the case when the values of $M$ 
in Eq.~\eqref{eq:Mt} are taken to be $M_1 = 2.5$ and $M_2 = 3.5$. The 
other parameter values are $\De_1 = 1$ and $\De_2 = 0.1$.} \end{figure}

Since each eigenvalue $\exp({-i\ep_{j}T})$ is a complex number of unit 
magnitude, it can be represented as a point on the unit circle. 
Figure~\ref{fig:qE} shows a plot of 
$\cos (\ep_j T)$ versus $\sin (\ep_j T)$. The points marked in red (shown more 
clearly in the inset) correspond to the eigenstates which are localized at one
of the corners. One such state is shown in Fig.~\ref{fig:corner_dyn}. These 
corner states are fourfold degenerate and lie close to $\sin (\ep_j T)= 0$. 
For this particular choice of parameters, these quasienergies lie at 
$\cos (\ep_j T)= -1$. They may also lie at $\cos (\ep_j T)= +1$ for a different 
set of parameters. For any choice of parameters, as long as we are in 
the higher-order topological sector, i.e., the diagonal winding number is 
nonzero, these corner states exist and are always separated from the 
bulk and edge states by a finite gap which is proportional to $\De_2$.

It is interesting to note that corner states can appear even when both $M_1$ 
and $M_2$ are larger than 2 as we can see from Fig.~\ref{fig:mirror_dyn}. Thus 
the periodic driving can generate corner states even when the instantaneous 
Hamiltonian does not have corner states at any time; as 
Fig.~\ref{fig:mirror_eq} shows, the time-independent Hamiltonian has 
no corner states if $M > 2$.

\section{Anisotropic model}
\label{sec:st_asym}

We now consider a variation of the model discussed so far by making the 
hopping amplitude $t_0$ anisotropic, i.e., taking the hoppings along $x$ and 
$y$ directions to be $t_x$ and $t_y$ respectively. In the case of the static 
system, the Hamiltonian in Eq.~\eqref{eq:con_ham} becomes 
\bea H (\bk) &=&\big(M ~+~ t_x ~\cos k_x ~+~ t_y ~\cos k_y\big) ~\tau^z
\otimes \si^0 \non \\
&&+ ~\De_1 ~(\sin k_x \tau^x\otimes\si^x ~+~ \sin k_y \tau^x\otimes
\si^y) \non \\
&&+ ~\De_2 ~(\cos k_x ~-~ \cos k_y) ~\tau^y\otimes ~\si^0. 
\label{eq:con_ham_asym} \eea
(In the anisotropic model, the coefficients of $\cos k_x$ and $\cos k_y$
in the last term in Eq.~\eqref{eq:con_ham_asym} could, in principle, 
be different, namely, $\De_{2x} \ne \De_{2y}$. However, we will assume
that they are equal, and we will call the coefficient $\De_2$ as before. This 
will allow us to define a diagonal winding number as before).
This Hamiltonian has the same symmetries as the one in Eq.~\eqref{eq:con_ham}
except for the ones which involve the $\mc{C}_4$ transformation.
Further, unlike the isotropic model, 
the energy spectrum does not remain the same if we interchange $k_x$ and $k_y$.

We can find the energy spectrum of Eq.~\eqref{eq:con_ham_asym} in the same
way as given in Eq.~\eqref{spec1}. We find that
\bea E(\bk) &=& \pm ~[ (M ~+~ t_x \cos k_x ~+~ t_y \cos k_y))^2 \non \\
&& ~~~~+~ \De_1^2 ~(\sin^2 k_x ~+~ \sin^2 k_y) \non \\
&& ~~~~+~ \De_2^2 ~(\cos k_x ~-~ \cos k_y)^2]^{1/2}. \label{spec2} \eea
This implies that the bulk gap can vanish only if (i) $M= \pm (t_x + t_y)$
or (ii) $M = \pm (t_x - t_y)$ and $\De_2 = 0$. These give the locations
of topological phase transitions.

\begin{figure}[htb]
\begin{center}
\sg[~Chern number]{\ig[height=1.76cm]{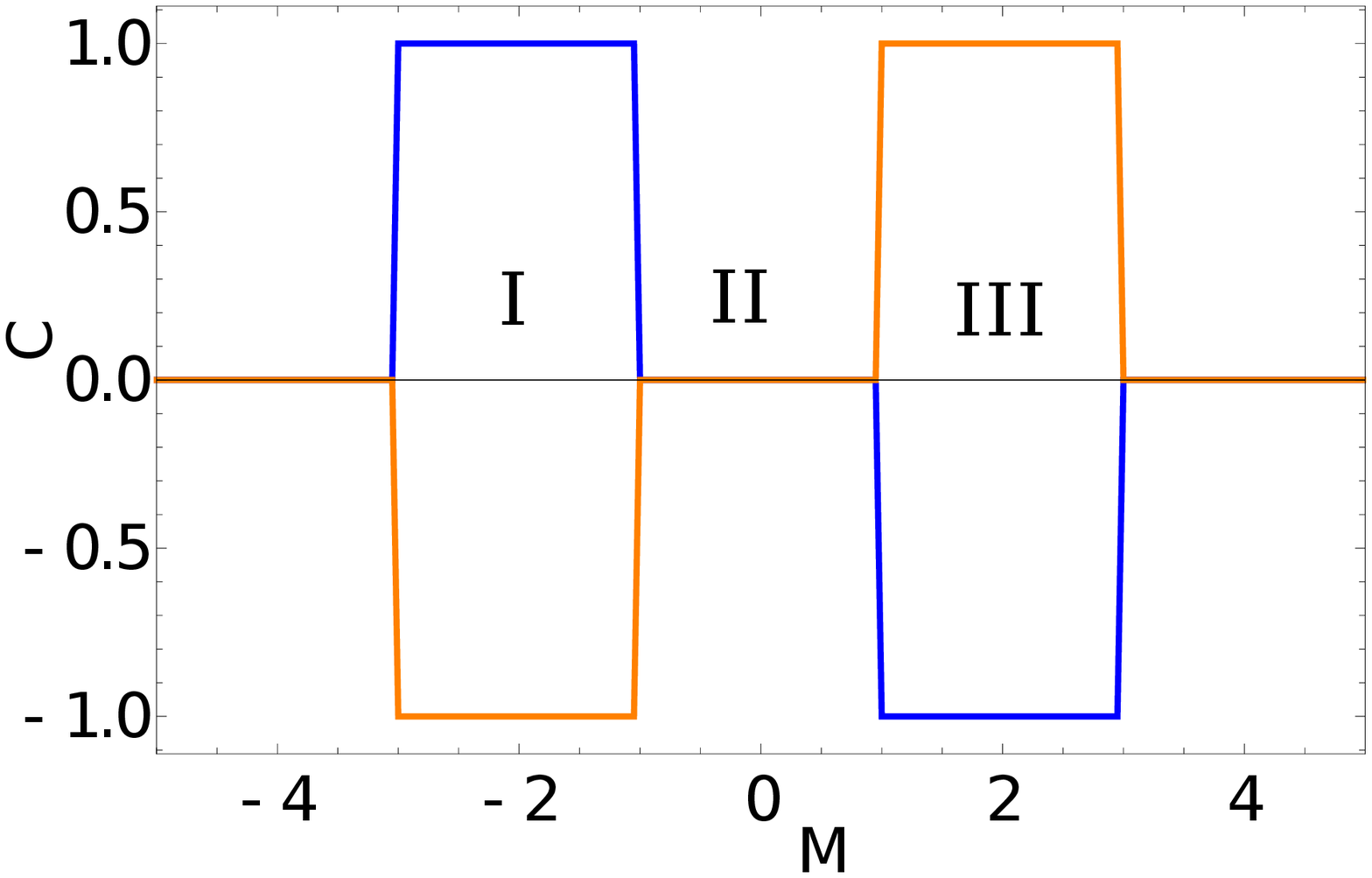}\label{fig:chern_asym}}
\sg[~Winding number]{\ig[height=1.76cm]{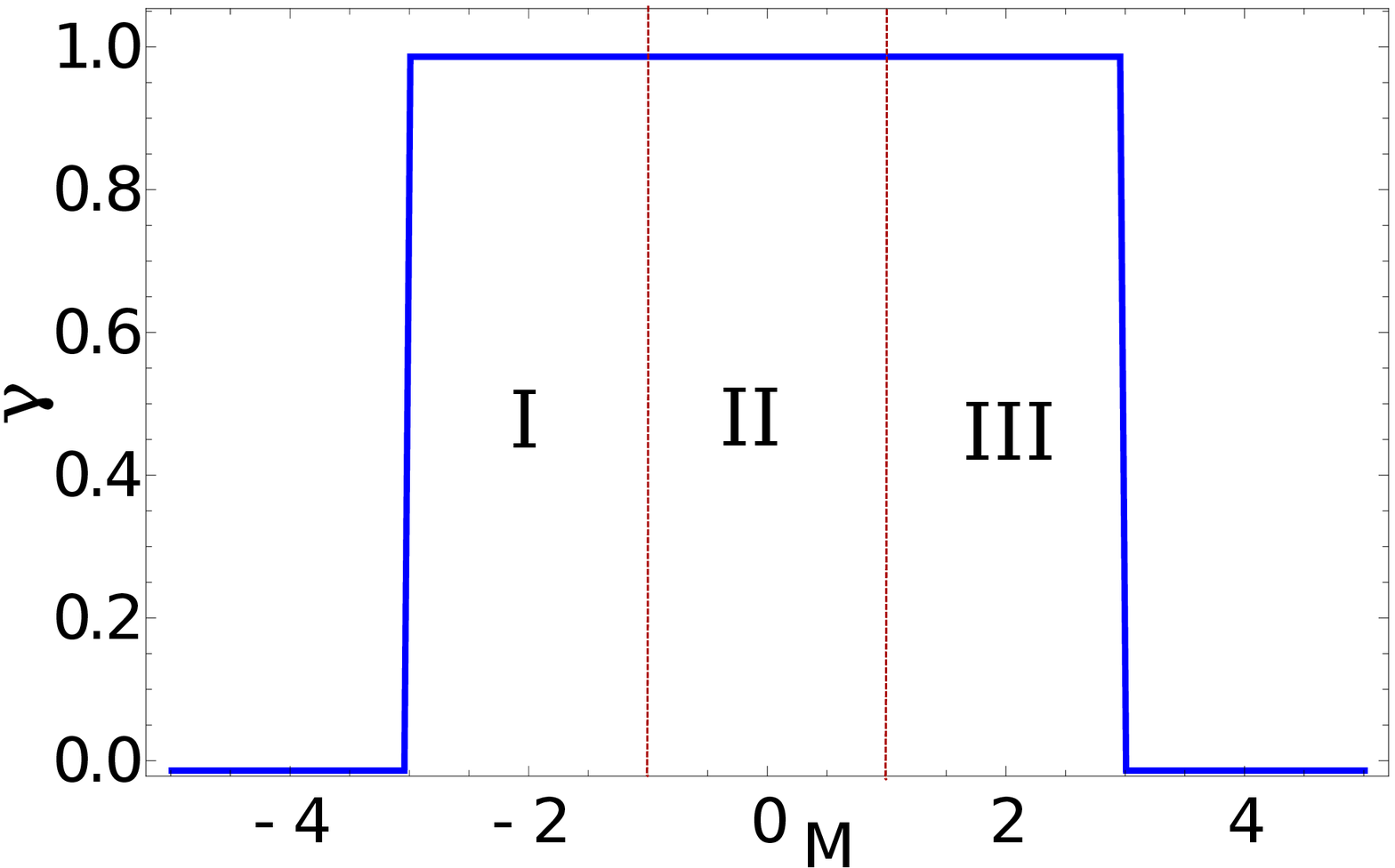}
\label{fig:mirror_asym}}
\sg[~$\ga=1$]{\ig[height=1.76cm]{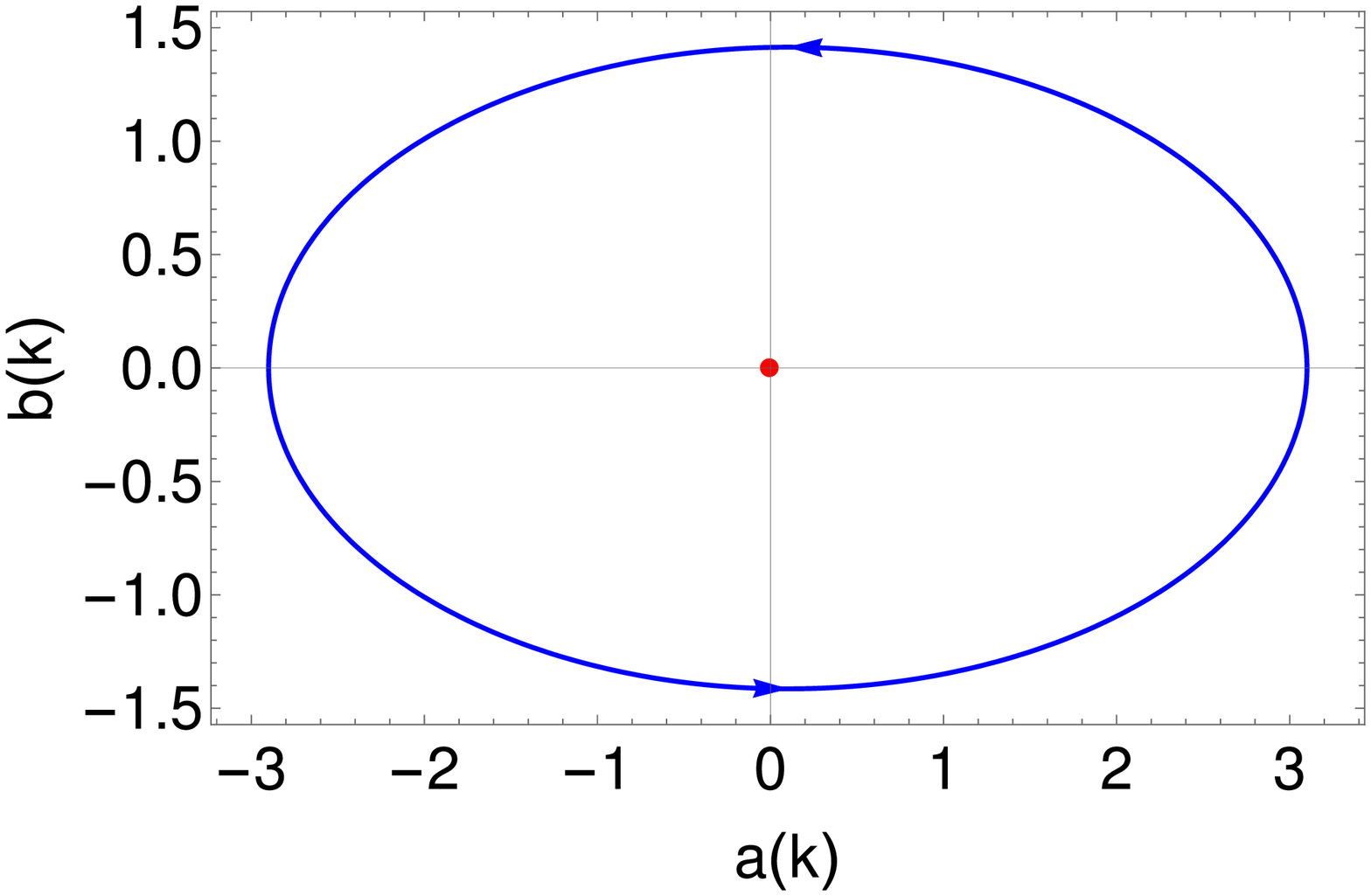}\label{fig:mir_win_anis_eq}}
\end{center}
\caption{Between the two topologically nontrivial regions $I$ and $III$, there
is a region $II$ where the Chern number of the system is zero. The width of 
this region is given by $2 ||t_x| - |t_y||$. In (a) we have taken $t_x = 1$ 
and $t_y = 2$ and have set $\De_2=0$ here. In (b) we show the diagonal winding 
number.
In (c) we show a parametric plot of $b(k_x)$ vs. $a(k_x)$ at $M=0.1$. As this 
curve goes around the origin once in the anti-clockwise direction, we conclude 
that $\ga=+1$.} \label{fig:inv_asym} \end{figure}

We can now follow the same procedure as outlined in Sec.~\ref{sec:st_bulk} to 
find the topological invariants, i.e., the Chern number and the diagonal 
winding number, corresponding to the cases with $\De_2 = 0$ and $\De_2 \neq 0$ 
respectively. Note that the transformations 
given in Eqs.~(\ref{eq:con_ham2}-\ref{abk}) which lead to a point in a 
two-dimensional plane and hence to a winding number continue to work in this
anisotropic system, as long as the system is gapped at all momenta. The 
phase diagram and the values of the topological invariants are shown in 
Fig.~\ref{fig:inv_asym}.

In the isotropic case, as discussed in Sec.~\ref{sec:st_bulk}, at $M=0$ there 
is a transition from one topologically nontrivial phase to another (regions 
labeled $A$ and $B$ in Fig.~\ref{fig:chern_eq}). However, on introducing an 
anisotropy by taking $t_x \neq t_y$, the two topologically nontrivial regions 
are separated by an intermediate region where the Chern number is zero. (The 
boundaries of all the regions can be found by finding the values of $M$ where 
the energy eigenvalues of Eq.~\eqref{eq:con_ham_asym}, with $\De_2 = 0$, vanish
at one the four time-reversal invariant momenta). For the case $t_x =1$ and 
$t_y =2$, we see in Fig.~\ref{fig:chern_asym} that the intermediate phase, 
labeled as $II$, lies in the region $|M|\leq 1$, whereas the phases labeled 
$I$ and $III$ are topologically nontrivial with Chern numbers $C = \pm 1$ 
respectively. The width $W_M$ of phase $II$ depends on the hoppings as 
$W_M = 2||t_x|-|t_y||$. 

\vspace*{.2cm}

On studying the Hamiltonian on a square lattice with $t_x \ne t_y$, we find 
that the edge states in the topological and nontopological phases behave very
differently from each other. In regions $I$ and $III$, the edge states are 
present along all the edges of the system and are topologically protected. 
However, in region $II$, the edge states exist only on the edges parallel to 
the $x$ direction (when $|t_x| < |t_y|$) and are not topologically protected,
even though they are gapless (if $\De_2 = 0$) and they lie inside the bulk gap.
This is consistent with the value of the Chern number which is zero in region 
$II$. If $|t_y| < |t_x|$, the edge states are found only on the edges parallel 
to the $y$ direction. The system can therefore be described as a {\it weak} 
topological insulator in region $II$ for $\De_2 = 0$~\cite{hasan,qi,fu}.

For $\De_2=0$, the different phases of the system can be distinguished from
each other by the Chern number as shown in Fig.~\ref{fig:chern_asym}. If 
$\De_2 \ne 0$, the Chern number is ill-defined in all the phases, but they 
can be distinguished from each other by the diagonal winding number (whose
value does not depend on $\De_2$). 
Figure~\ref{fig:mirror_asym} shows the winding number $\ga$ for the 
anisotropic system with $\De_2 = 0.3$. We see that $\ga = 1$ for $|M|\leq 3$.
The winding number does not differentiate among the phases $I$, $II$ 
and $III$ which are topologically different from one another as shown by the 
Chern number. We thus see that the regions of nonzero Chern and 
diagonal winding numbers are not identical in the anisotropic model.

\begin{figure}[htb]
\begin{center}
\sg[~Region $I$: $M = -2$]{\ig[height=3.1cm]{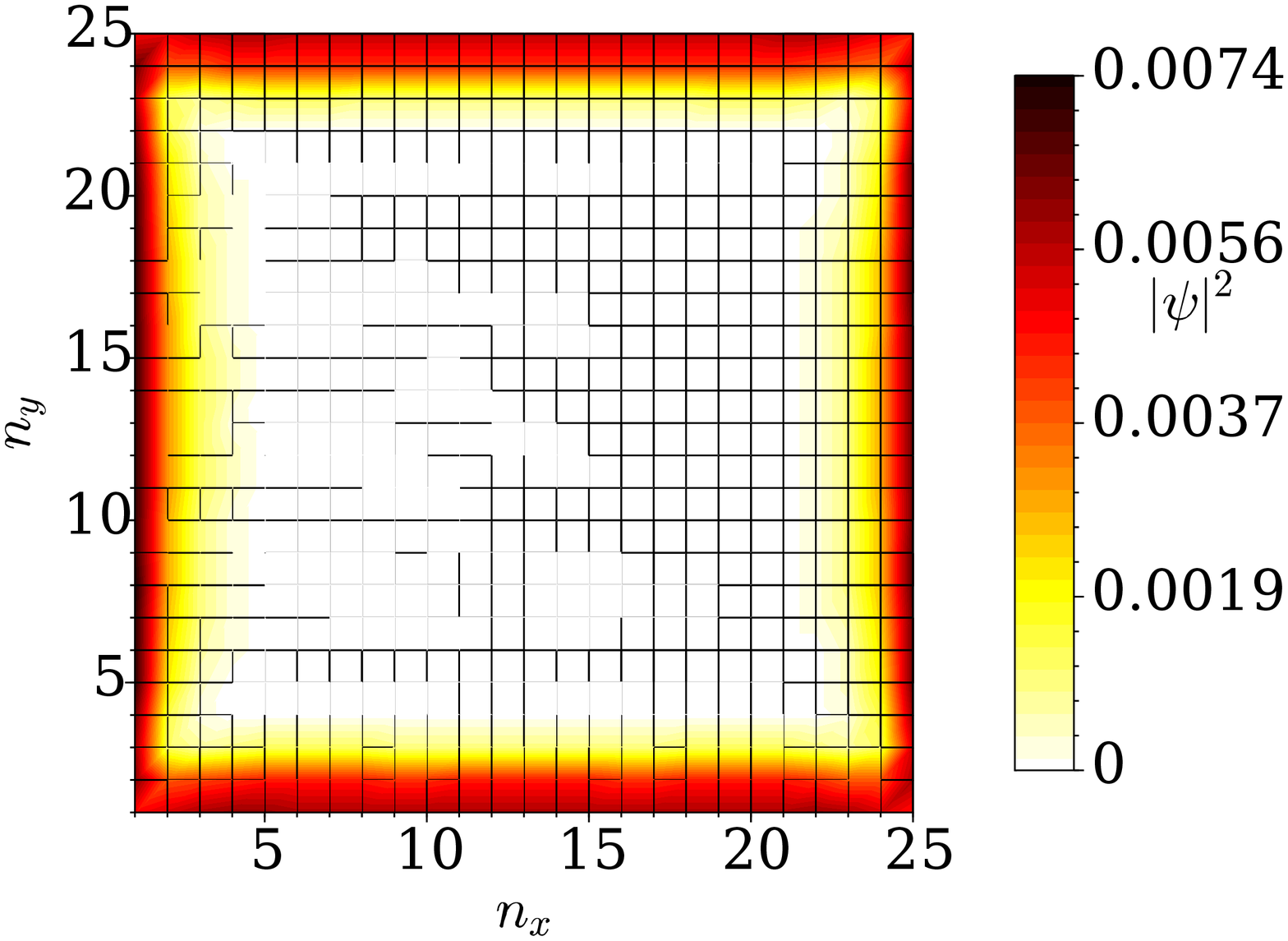}}
\sg[~Region $II$: $M = 0.5$]{\ig[height=3.1cm]{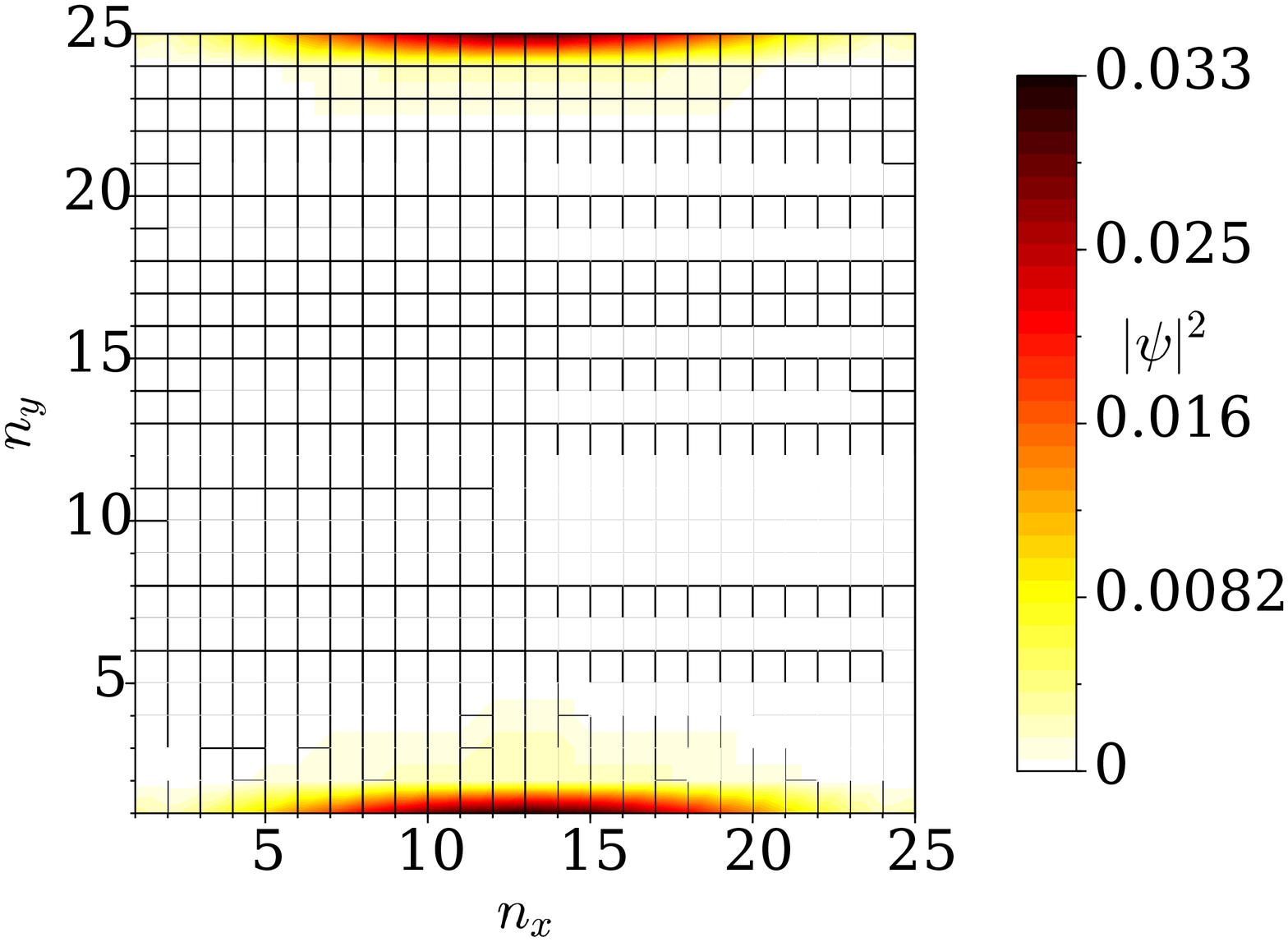}}
\end{center}
\caption{Edge states for phases corresponding to regions $I$ and $II$ for a
system with the same parameters as in Fig.~\ref{fig:inv_asym}; the energies
of the states are $0.73$ and $0.387$ in units of $t_x$ in figures (a) and (b) 
respectively. In region $I$, since the Chern number is nonzero, the edge 
states exist along all edges of the sample. However, in region $II$ where the 
Chern number is zero, the edge states exist only along the edges parallel to 
the $x$ direction. We have taken $t_x =1$, $t_y = 2$, $\De_1 = 1$ and 
$\De_2 = 0$.} \label{fig:edge_asym} \end{figure}

\begin{figure}[htb]
\begin{center}
\sg[~Energy]{\ig[width=4.cm]{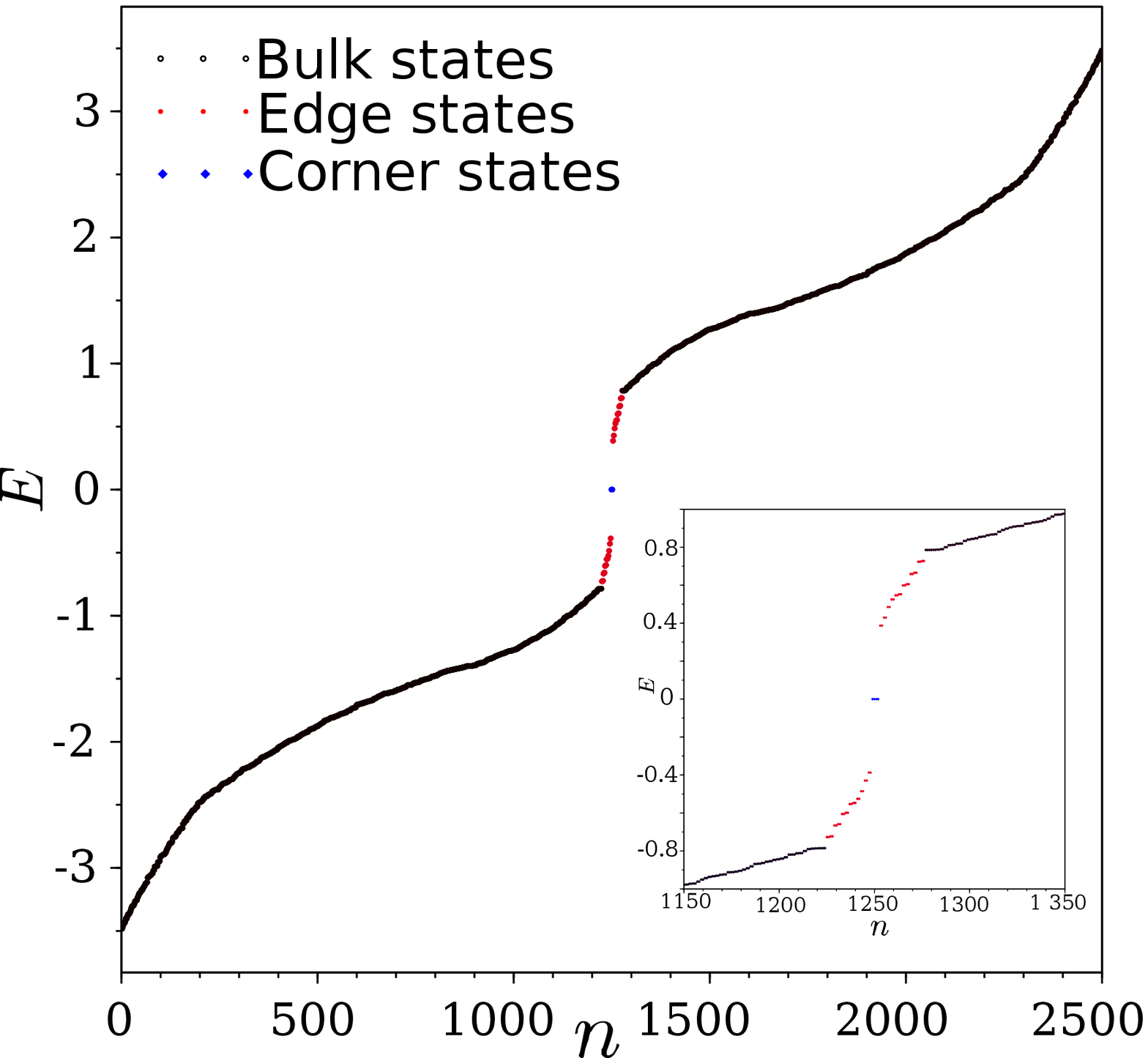}}
\hspace{0.3cm}
\sg[~Corner state]{\ig[width=4.cm]{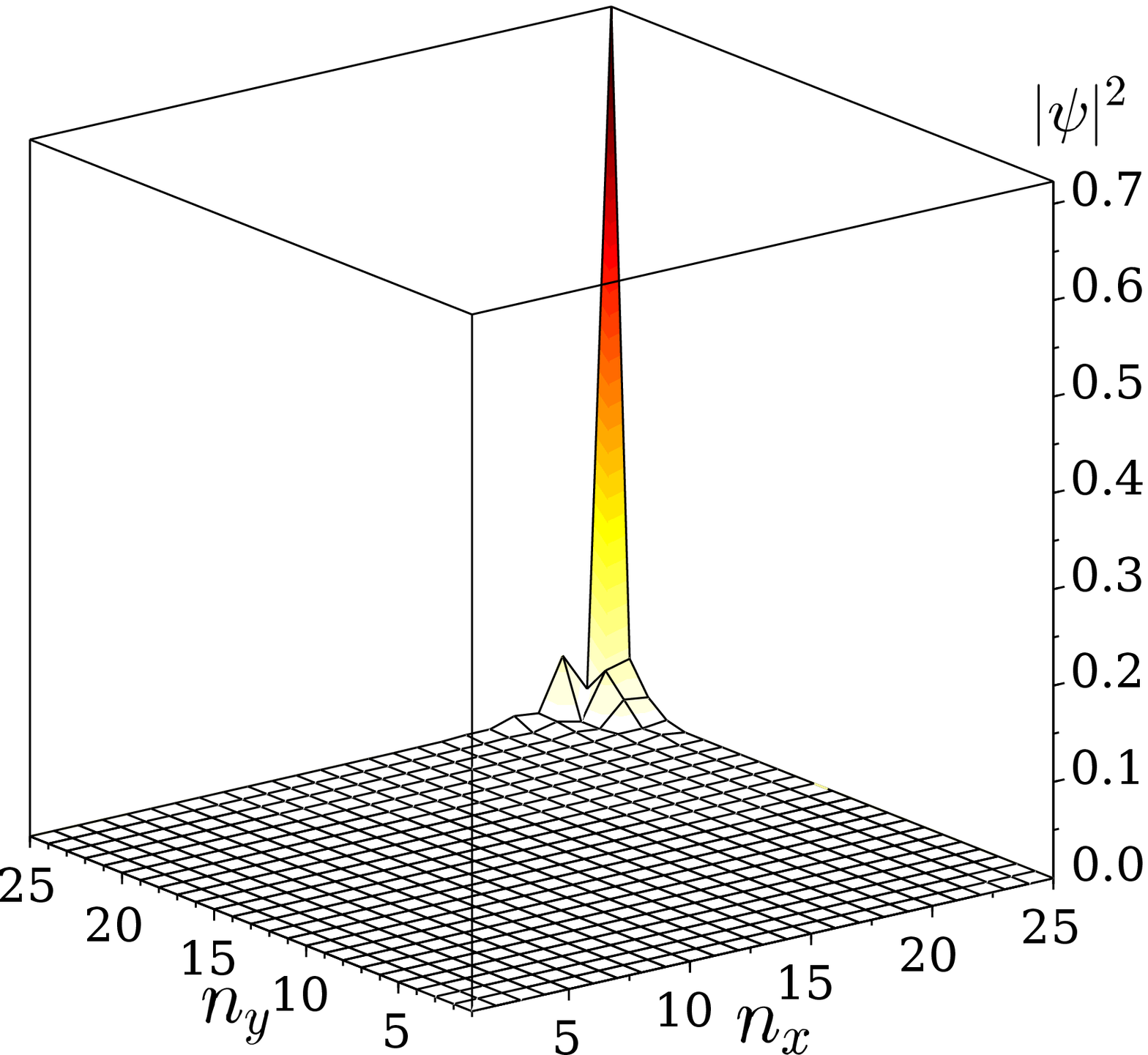}}
\end{center}
\caption{(a) Energy eigenvalues and (b) a corner state as found for a 
$25\times25$ square lattice in region $II$ where the Chern number is zero in 
Fig.~\ref{fig:chern_asym}. The corner states are at zero energy and are 
separated from the rest of the states by an energy gap of about $\De_2$. This 
is clear from the inset in (a). The state shown in (b) is one of the four 
corner states at zero energy. The edge states in this region lie on only 
two of the edges and there are robust corner states. The edge states in this 
region are found to exist only along the edges parallel to the $x$ direction.
We have taken $M = 0.5$, $t_x =1$, $t_y = 2$, $\De_1 = 1$ and $\De_2 = 0.3$.}
\label{fig:corner_asym} \end{figure}

The corner states are obtained by diagonalizing the Hamiltonian in real space 
on a square lattice. The expression for this Hamiltonian is similar to 
Eq.~\eqref{eq:Ham_sq} with the hoppings along $x$ and $y$ directions
being $t_x$ and $t_y$ respectively. Figure~\ref{fig:edge_asym} shows the edge 
states in regions $I$ and $II$ for $\De_2 = 0$. We see that in region $I$, the
edge states lie on all the edges while in region $II$, they only lie on the 
edges parallel to the $x$ direction (for $|t_x| < |t_y|$). This is consistent 
with the Chern numbers in these two regions. Figure~\ref{fig:corner_asym} 
shows the energy eigenvalues (the corner states lie at zero energy) and 
a corner state in region $II$ for $\De_2 = 0.3$. 


We now study the effect of periodically driving the system by varying the 
parameter $M$ between two values $M_1$ and $M_2$ both of which lie in the 
nontopological regime, i.e., in region $II$ of Fig.~\ref{fig:inv_asym}. 
The values of $T$ where topological transitions occur depend on the values 
of $M_1$ and $M_2$ chosen in Eq.~\eqref{eq:Mt}. 

\begin{figure}[htb]
\sg[~Chern number]{\ig[height=2.9cm]{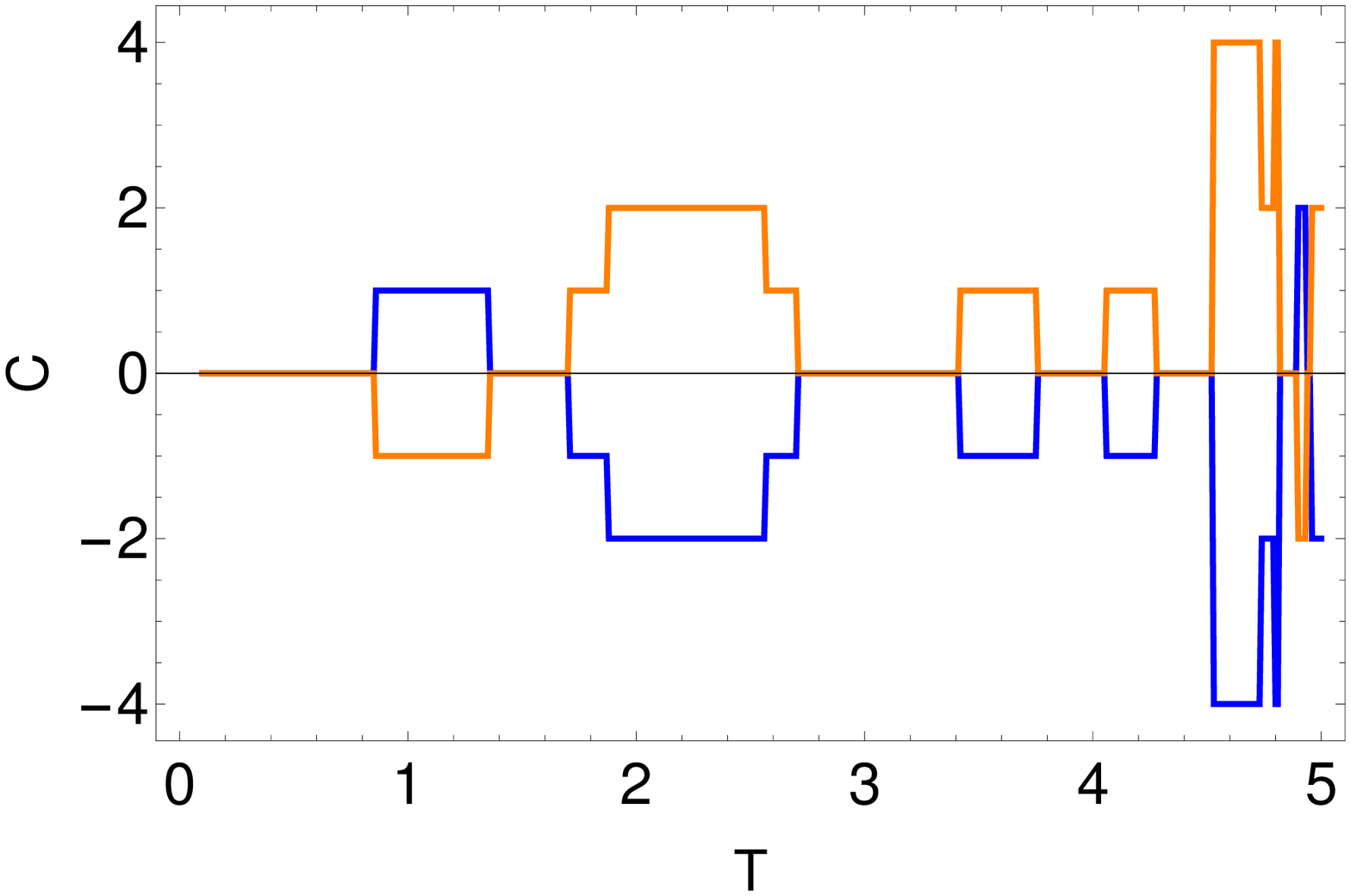}} 
\hspace*{.2cm} \sg[~Winding number]{\ig[height=2.9cm]{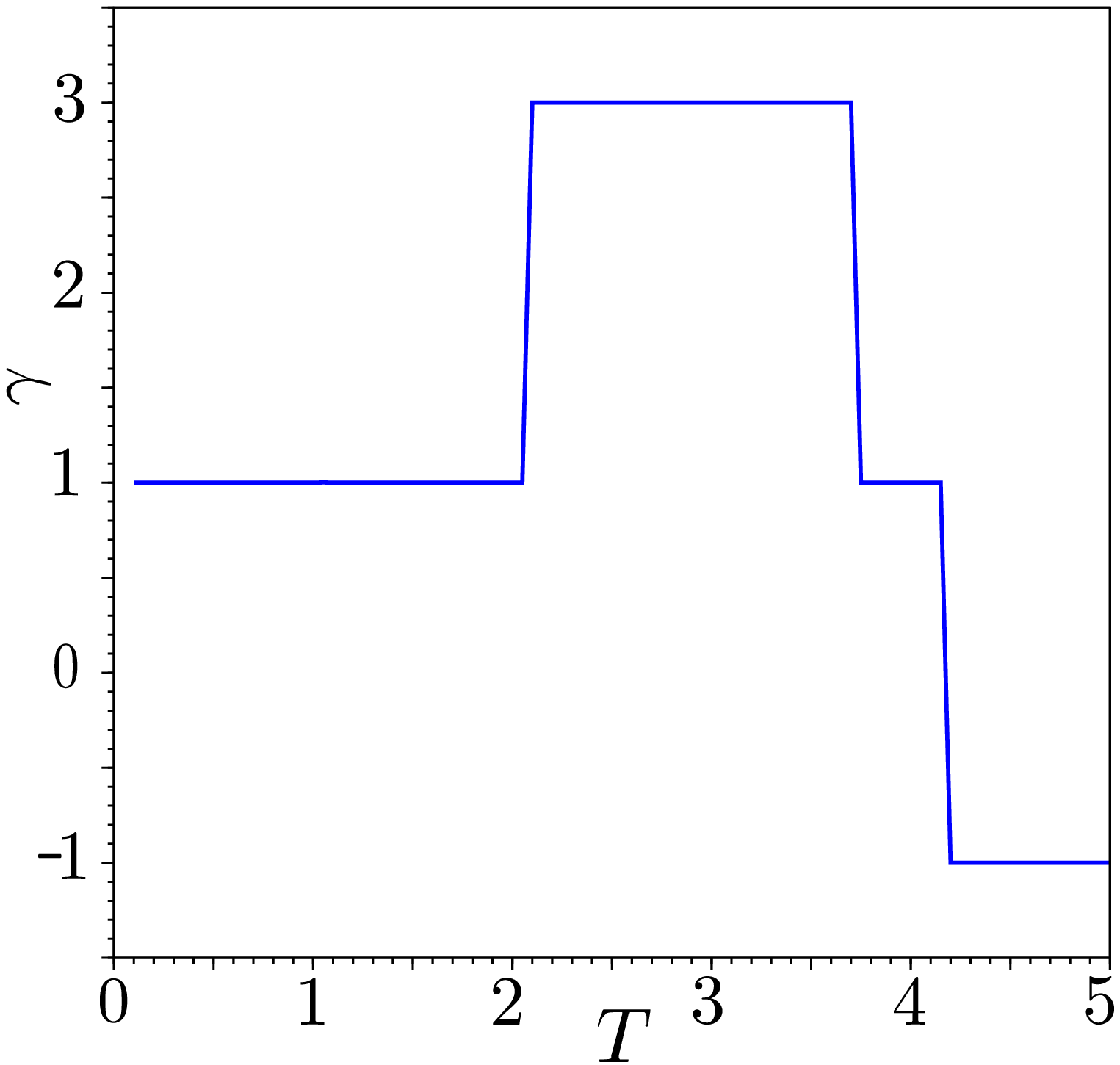}}
\caption{Topological invariants calculated from the Floquet operator of the 
Hamiltonian in Eq.~\eqref{eq:con_ham_asym} as a function of $T$ for 
$M_1=-0.9, M_2= - 0.45, ~\De_1=1, ~t_x=1$ and $t_y=2$. In (a) we have set 
$\De_2=0$ and plotted the Chern number calculated from the Floquet eigenstates 
showing jumps at certain values of $T$ indicating topological phase 
transitions. In (b) we show the diagonal winding number; this shows jumps at 
certain values of $T$ which are consistent with those obtained from the 
method described in Appendix A.} \label{fig:asym_con_dyn} \end{figure}

In Fig.~\ref{fig:asym_con_dyn} (a), the Chern number is calculated for the 
anisotropic Hamiltonian in Eq.~\eqref{eq:con_ham_asym} for $\De_2=0$. 
The jumps in the Chern number at some specific values of the parameter $T$
signify topological transitions and can be understood as discussed earlier. 
We note that $U_F (\bk)= \pm ~\mathbb{I}_4$ at these points and the 
magnitude of the Chern number can exceed 1 unlike the static system.
It turns out that the jumps in the Chern number occurring for $T < 4.4$
are due to contributions from the time-reversal invariant momenta, while
the jumps occurring for $T > 4.4$ are due to contributions from other momenta 
which have no special symmetries. This is explained in Appendices A and B 
respectively. We also see some sharp fluctuations in the Chern number at 
$T = 4.81$ and $4.91$; the reasons for these are explained in Appendix B.

We can compute the diagonal winding number $\ga$ from the wave functions of 
the quasienergy states of the Floquet operator; the value of $\ga$ does 
not depend on $\De_2$. The winding number also jumps at some specific values 
of $T$ signifying second-order topological transitions. 
\vspace*{.2cm}

\begin{figure}[H]
\begin{center}
\sg[~$\ga = 1$ ]{\ig[width=4.25cm]{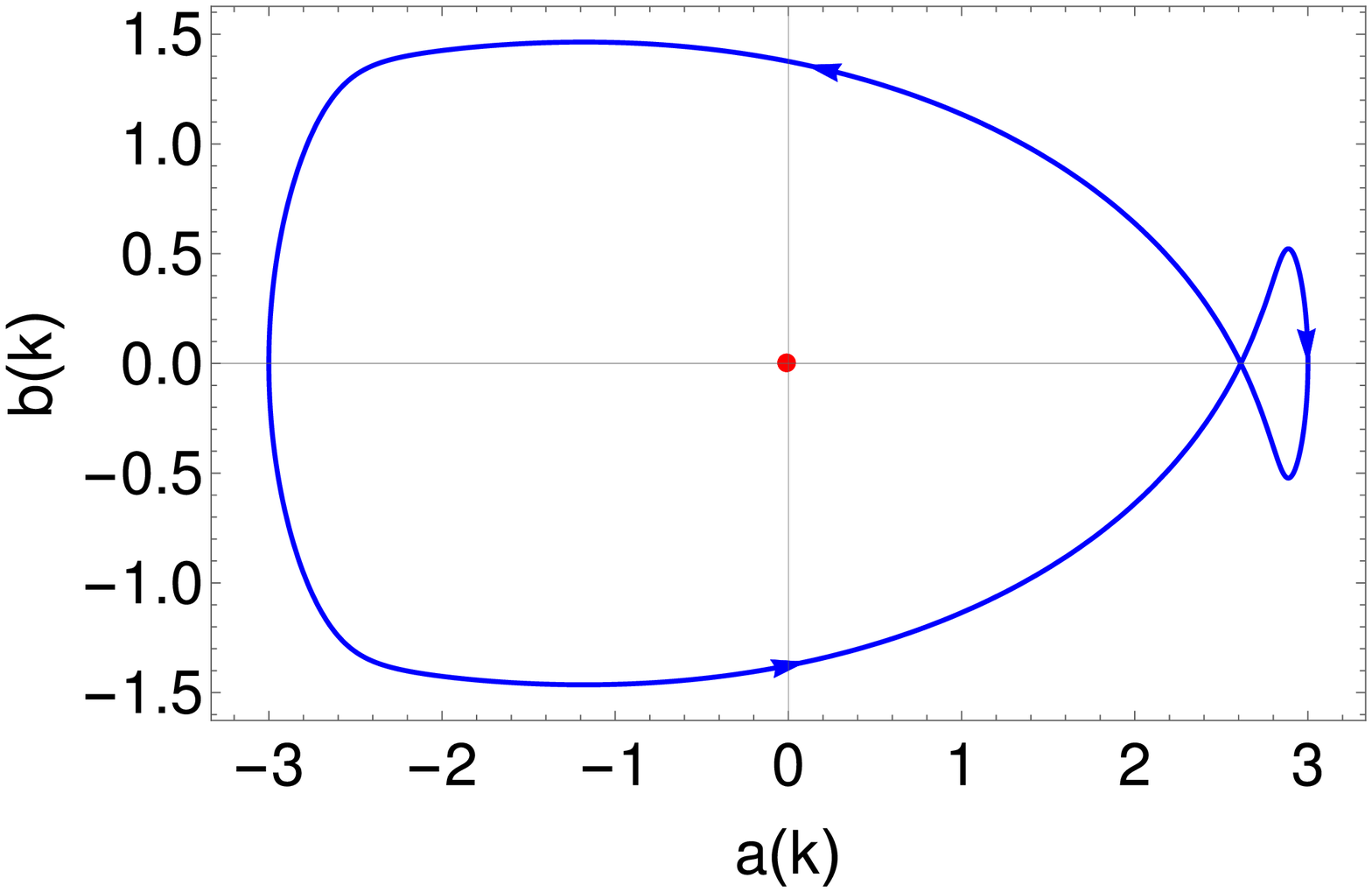}\label{fig:mir_win_anis_dyn1}}
\sg[~$\ga = 3$ ]{\ig[width=4.25cm]{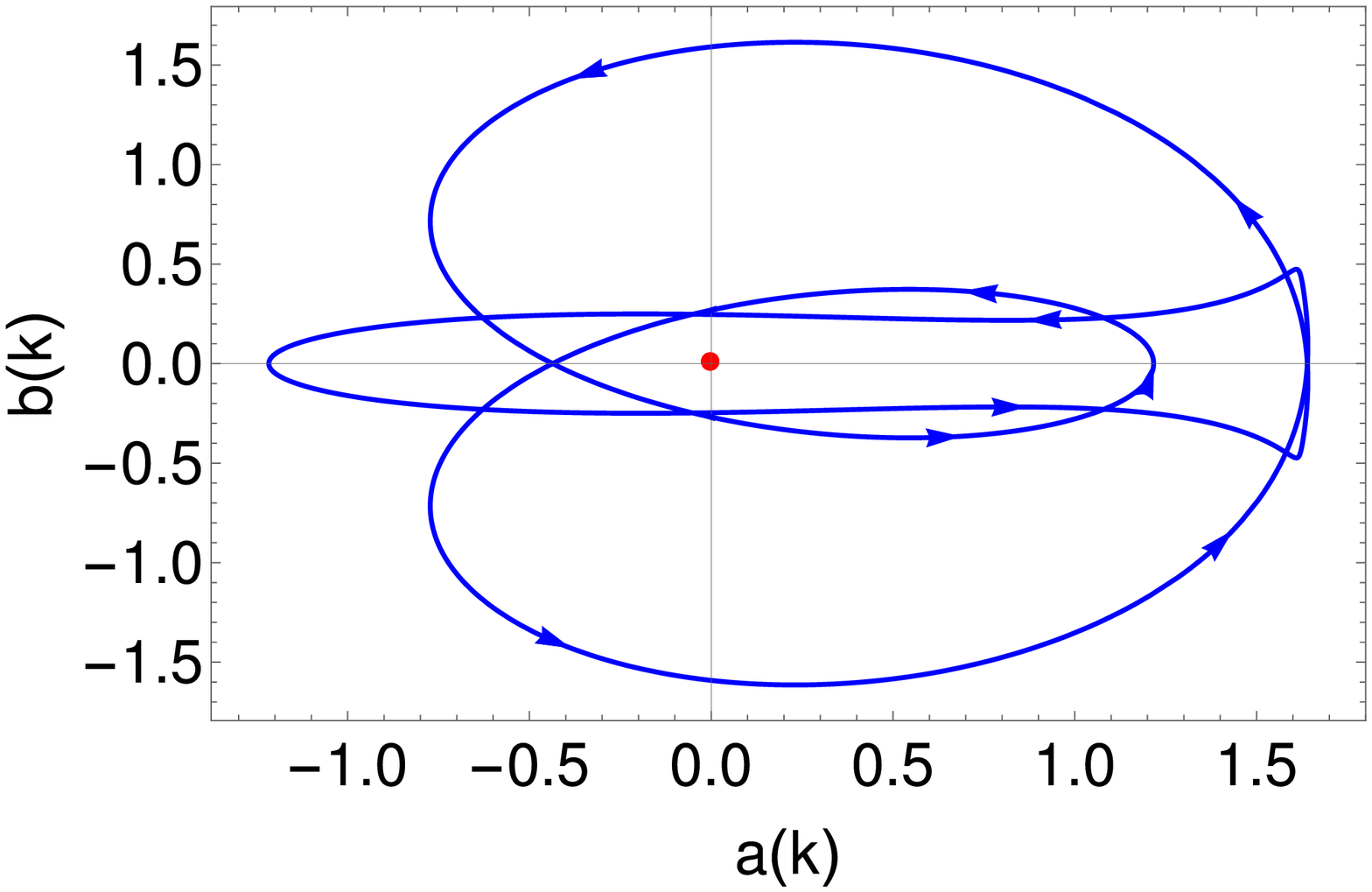}\label{fig:mir_win_anis_dyn3}}	
\end{center}	
\caption{Floquet diagonal winding numbers for periodically driven anisotropic 
model for (a) $T = 1$ and (b) $T=3$. In both cases we have taken $M_1 = -0.9$,
$M_2 = 0.9$, $\De_1 = 1$, $t_x = 1$ and $t_y = 2$. The number of times these 
parametric curves of $b(k_x)$ vs. $a(k_x)$ wind around the origin is the 
winding number.} \label{fig:asym_mirror_lat} \end{figure} 

\begin{widetext}
\begin{figure}[H]
\begin{center}
\sg[~Quasienergies]{\ig[height=3.8cm]{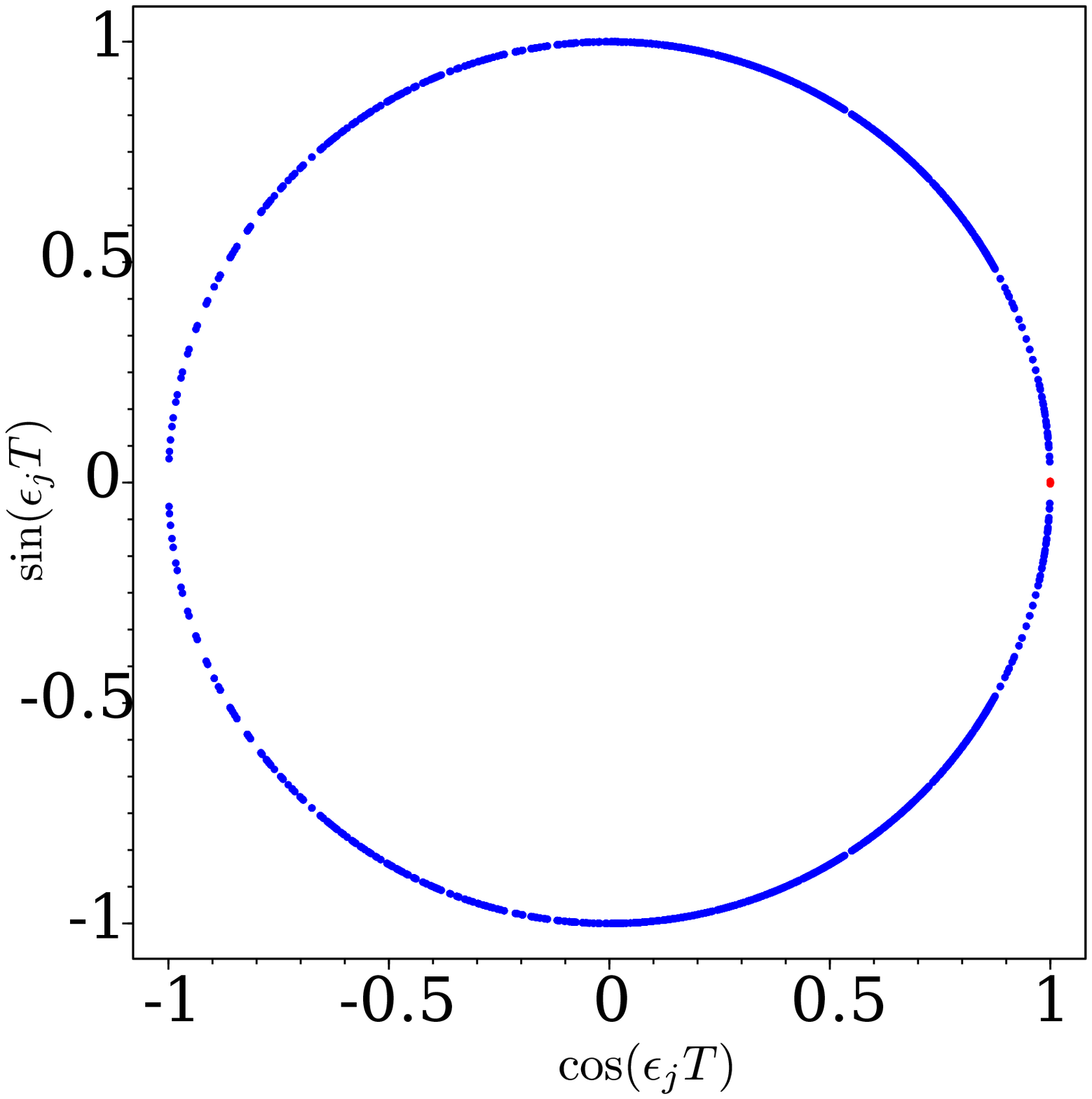}}\label{fig:asym_dy_qE}
\sg[~Corner State 1]{\ig[height=4cm]{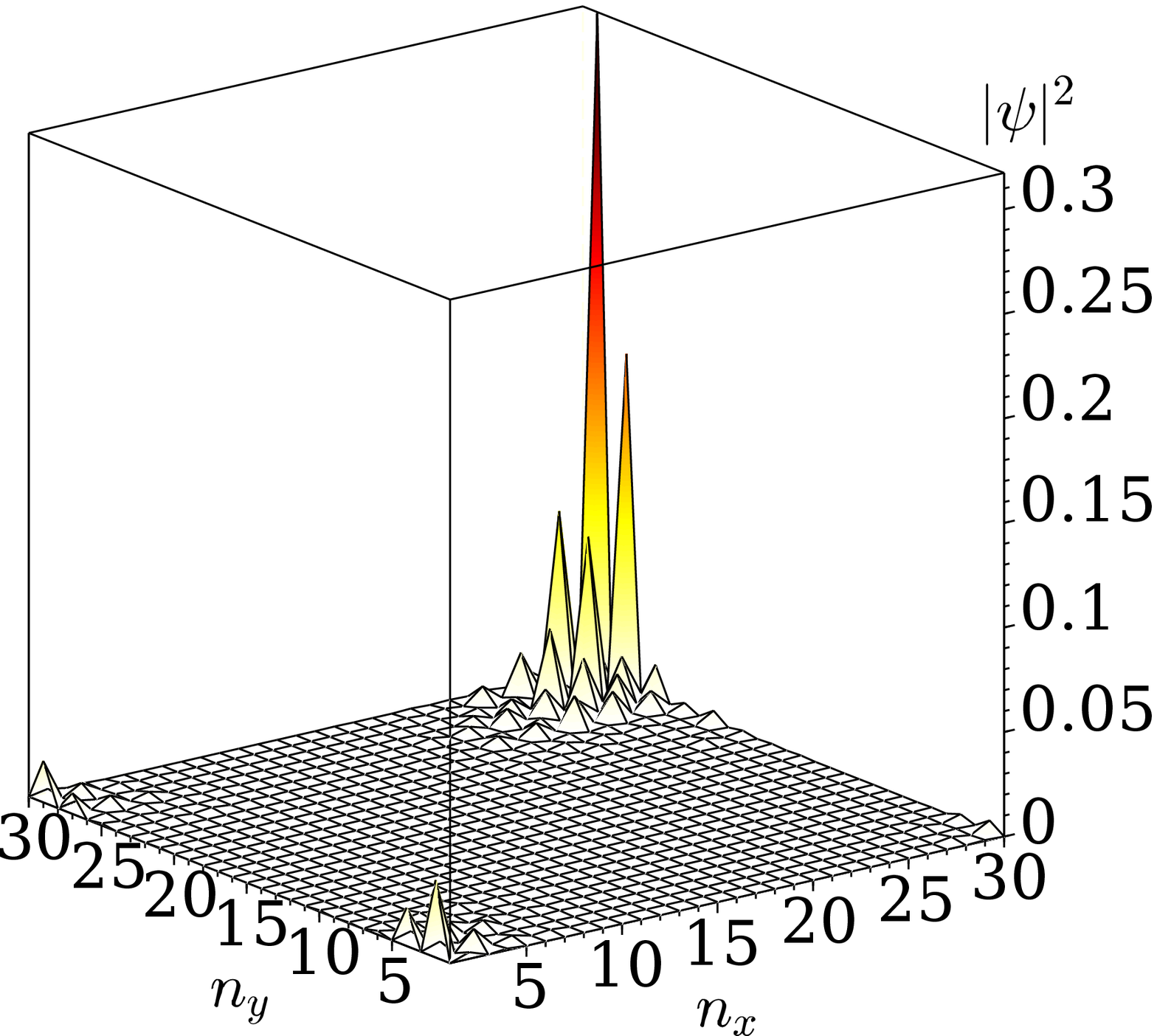}\label{fig:asym_dy_C1}}
\sg[~Corner State 2]{\ig[height=4cm]{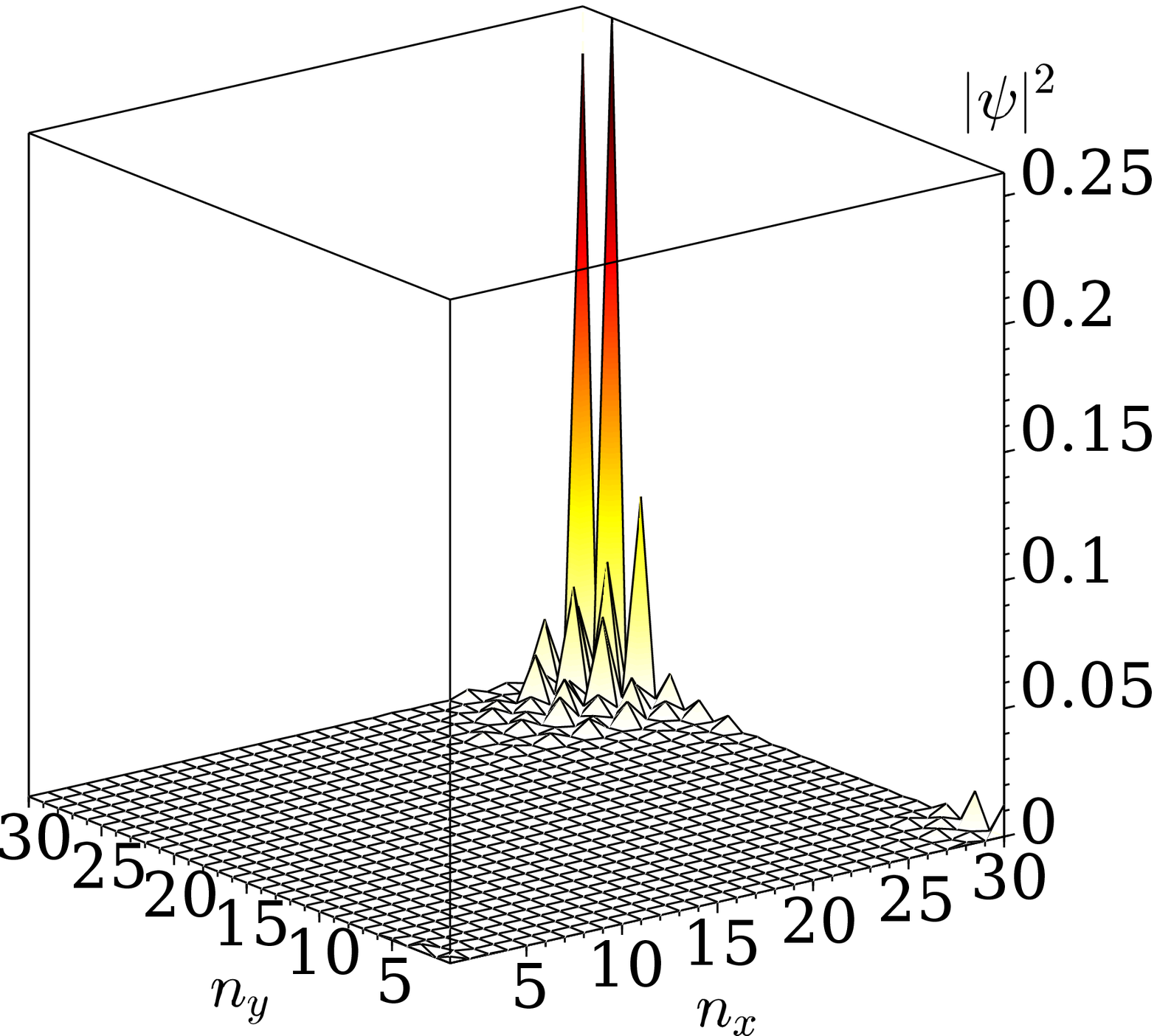}\label{fig:asym_dy_C2}}	
\sg[~Corner State 3]{\ig[height=4cm]{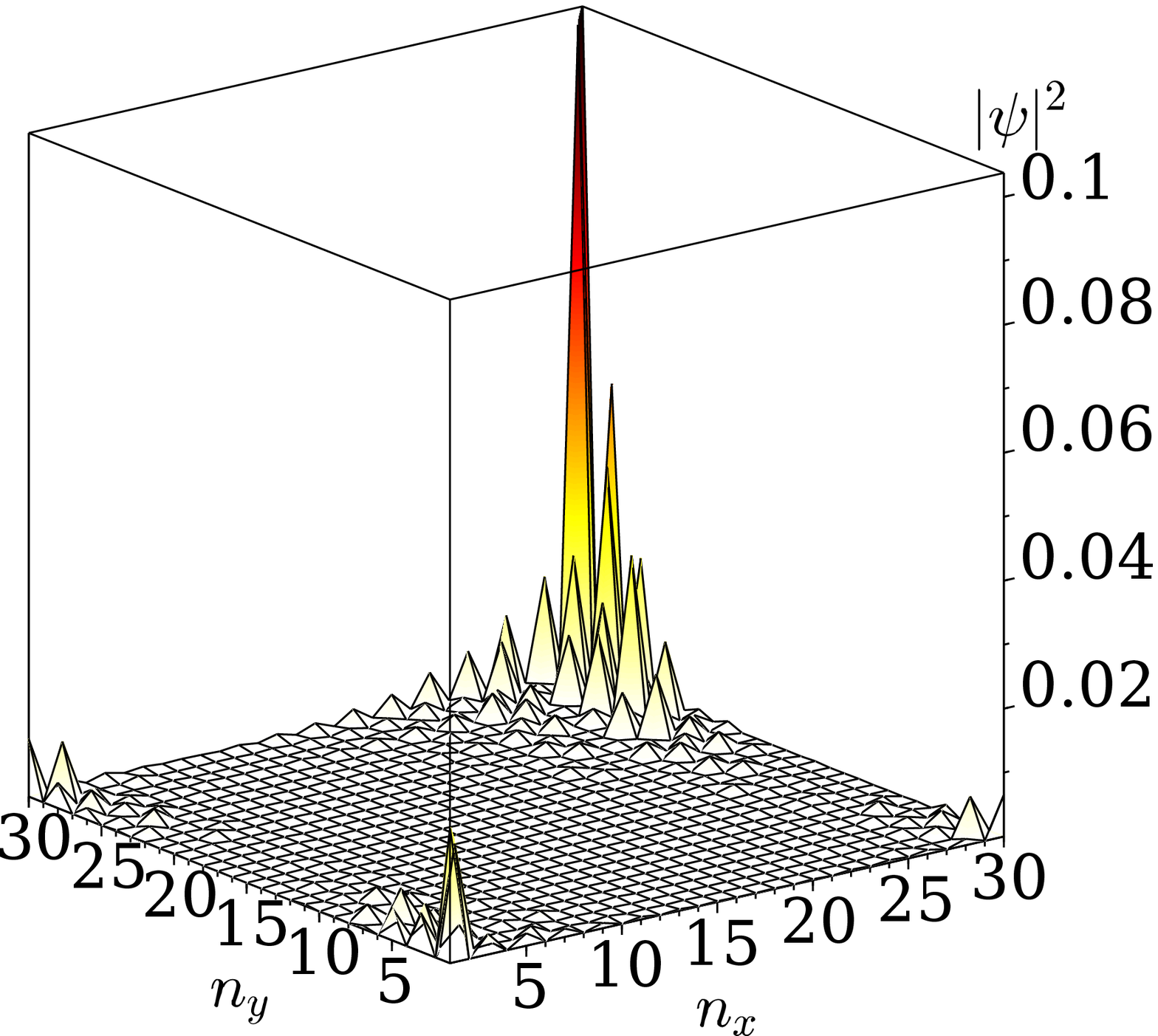}\label{fig:asym_dy_C3}}
\end{center}	
\caption{Multiple corner states on a square of size $30 \times 30$ sites with 
periodic driving. We have taken $M_1 = -0.9$, $M_2 = 0.9$, $t_x = 1$, $t_y = 
2$, $\De_1 = 1$, $\De_2 = 0.3$ and $T = 3$. In (a) we have plotted the 
quasienergies of the bulk, edge and corner states calculated from the Floquet 
operator. There are 12 corner states close to zero energy which are 
degenerate in the thermodynamic limit. By taking appropriate superpositions 
among these states, we find three states at each corner, as shown in (b), (c) 
and (d). The corner states decay exponentially into the bulk.} 
\label{fig:asym_qenergy_lat} \end{figure} 
\end{widetext}

Finally we consider a finite-sized square with $30 \times 30$ sites and 
periodically drive the anisotropic lattice Hamiltonian. 
From the momentum-space calculations, we find regions with $\ga = 1$ and 3, 
as shown in Figs.~\ref{fig:mir_win_anis_dyn1} and \ref{fig:mir_win_anis_dyn3}.
For the lattice calculations, we will work with the driving period, i.e., 
$T = 3$, and demonstrate that the diagonal winding number 3 indeed counts the 
number of corner states. The corner states appear in 
this system only for those parameter values for which the winding 
number of the Hamiltonian in Eq.~\eqref{eq:con_ham_asym} is nonzero. For 
low-frequency driving, i.e., when the time period of driving is large compared
to the timescale $1/t_0$, we sometimes find more than 4 corner states in the 
system which become degenerate at zero quasienergy in the thermodynamic limit. 
Figure~\ref{fig:asym_qenergy_lat} shows that for some values of the parameters,
there are a total 12 corner states (3 at each corner), all lying at zero 
quasienergy in the thermodynamic limit. All the three states shown at 
corner 3 have the eigenvalue -1 for the operator $\mc{S}_2 = \tau^x \si^z$.

\onecolumngrid


\section{Discussion}
\label{sec:disc}

We first summarize the results obtained in this work. We considered a variant 
of the BHZ model in which there is an additional term (with a coefficient 
$\De_2$) which breaks the $\mc{C}_4 $ and $\mc{T}$ symmetries separately but 
is invariant under their product. We first considered an isotropic model
and studied how its properties change as a parameter $M$ is varied.
The system can be characterized by two topological invariants called the 
Chern number and diagonal winding number.
The Chern number can be calculated only when the parameter $\De_2 = 0$,
while the winding number can be calculated for any value of $\De_2$ and it 
does not depend on the value of $\De_2$. When $\De_2 = 0$, there are no 
corner states. If the Chern number is nonzero, there are gapless edge states;
this system is then in a usual TI phase (called first-order). When 
$\De_2 \ne 0$ and the winding number is nonzero, there are gapped edge states
and gapless states which are localized at the corners of a finite-sized square 
lattice; the system is then in a second-order TI phase. We have shown that 
corner states have an interesting structure, namely, they are eigenstates of 
the operator $\tau^x \si^z$. We then studied the effect of periodic driving 
of the parameter $M$. We find that 
the driving can generate corner states in certain ranges of parameters where 
the corresponding time-independent Hamiltonian has no such states. We have
shown that the number of corner states is given by the winding number.

Next, we have studied an anisotropic version of the model
where the hopping amplitudes in the $x$ and $y$ directions are different.
Once again, we can define a Chern number if $\De_2 = 0$ and a diagonal 
winding number for any value of $\De_2$. An interesting feature of this model 
is that when $\De_2 = 0$, there is a nontopological phase in which the Chern 
number is zero and there are states localized along only two out of the four 
edges of a finite-sized square. When $\De_2 \ne 0$ and the winding 
number is nonzero, gapless corner states appear. We again find that periodic 
driving of $M$ in this anisotropic model can generate corner states and their
number is given by the winding number. 

For the periodically driven models (both isotropic and anisotropic),
we sometimes find that there can be more than one state localized at each
corner of the system. We have shown that the locations of the jumps in the 
Chern number and diagonal winding number can be understood as the points where 
the Floquet operator becomes proportional to the identity matrix at one of the
time-reversal invariant momenta and, in the case of the Chern number, 
at other momenta where there are no special symmetries; these jumps signify 
topological transitions. Although there has been no explicit discussion
in the earlier literature of momenta with no special symmetries which can 
lead to jumps in the Chern number, the role of such momenta has been alluded 
to in Refs.~\onlinecite{top25}, \onlinecite{top26} and \onlinecite{bomantara}.

We have shown that the diagonal winding number can also predict the number of 
states at each corner for the periodically driven system. 
A possible direction for future study
would be to look for a topological invariant which can predict the number of 
corner states with eigenvalues of $\tau^x \si^z = \pm 1$ separately. For 
instance, an interesting topological invariant to study would be the winding 
number of the one-dimensional states which are localized at one particular 
edge of an infinitely long ribbon with a finite width; these states would be 
labeled by a momentum $k_x$ (as shown in Fig.~\ref{fig:E_kx}), and an 
expression similar to Eq.~\eqref{eq_mirror_winding2} could then be used to 
calculate the winding number. At first sight, such a numerical calculation 
appears challenging because a finite width allows hybridization between the 
states at the opposite edges and we would therefore generally find states 
which are superpositions of states localized at the two edges. However, if 
these can be projected to one of the edges in a way which varies smoothly 
with $k_x$, it may be possible to calculate the corresponding winding number. 
This may provide a way of understanding the number of corner states.

\vspace{.8cm}
\centerline{\bf Acknowledgments}
\vspace{.5cm}
R.S. and A.D. thank Adhip Agarwala for useful discussions. A.D. acknowledges 
funding from SERB, DST, India for NPDF Research Grant No. PDF/2016/001482. 
D.S. thanks DST, India for Project No. SR/S2/JCB-44/2010 for financial support.

\vspace{.5cm}

\appendix

\section{Contributions to Chern and diagonal winding numbers from 
time-reversal invariant momenta} \label{app:k_points}

The jumps in the Chern number shown in Figs.~\ref{fig:chern_dyn} and
\ref{fig:asym_con_dyn} (a) can be understood by studying the 
time-evolution operators at some specific points in the Brillouin zone. 
These are the points where
\beq U_F (\bk) ~=~ \pm ~\mathbb{I}_4. \label{eq:Uid} \eeq 
First, let us study the anisotropic model and $\De_2 = 0$. Consider the 
Hamiltonian at the
momenta $\bk_{+} = (0,0)$ and $\bk_{-}=(\pi,\pi)$. At these two momenta the
$\De_1$ and $\De_2$ terms vanish. Therefore the Hamiltonians in the 
two halves of the cycle, $H_1$ and $H_2$ commute with each other and the 
time-evolution operator given in Eq.~\eqref{eq:U_lat} can be simplified to
\bea U_F (\bk_{\pm}) ~=~ e^{-i(M_1+ M_2 \pm 4t_0) (T/2) \tau^z}. 
\label{eq:U0pi} \eea
This gives the condition in Eq.~\eqref{eq:Uid} if
\beq (M_1 ~+~ M_2 ~\pm~ 4t_0) ~\frac{T}{2} ~=~ n\pi, \label{npi1} \eeq
where $n$ is an integer. For all our calculations in the isotropic case we 
have set $t_0 = 1$, $M_1 = 2.5$ and $M_2=3.5$. For $\bk_+$, the values of $T$ 
obtained from Eq.~\eqref{npi1} are given by $n\pi/5$
Similarly, for $\bk_-$, $T$ is found to be $n\pi$. 
These are precisely the points in Fig.~\ref{fig:chern_dyn} where the
Chern number jumps; all the jumps are by $\pm 1$ since there is a 
contribution from either $\bk_+$ or $\bk_-$ but not both. 

Similarly, the points $\bk_1 = (0, \pi)$ and $\bk_2 = (\pi,0)$ also contribute. 
In this case Eq.~\eqref{eq:U_lat} reduces to 
\bea U_F (\bk_{1,2}) ~=~ e^{-i(M_1+M_2) (T/2) \tau^z}, \label{eq:U2} \eea
which satisfies the condition in Eq.~\eqref{eq:Uid} if
\beq (M_1 ~+~ M_2) ~\frac{T}{2} ~=~ n\pi, \label{npi2} \eeq
where $n$ is an integer. This implies that there should be jumps in the Chern 
number at $T = n\pi/3$. Note that the contributions at these values of $T$ 
come from both $\bk_1$ and $\bk_2$. This explains the jumps of $\pm 2$ in the 
Chern number at these values of $T$ in Fig.~\ref{fig:chern_dyn}.

A similar analysis can be carried out for the anisotropic model where 
$t_x \ne t_y$. We then find that Eq.~\eqref{npi1} gets modified to 
\beq [M_1 ~+~ M_2 ~\pm~ 2(t_x + t_y)] ~\frac{T}{2} ~=~ n\pi, \label{npi3} \eeq
while Eq.~\eqref{npi2} changes to
\beq [M_1 ~+~ M_2 ~\pm~ 2(t_x - t_y)] ~\frac{T}{2} ~=~ n\pi. \label{npi4} \eeq
Equations~(\ref{npi3}-\ref{npi4}) explain the locations of the jumps in the 
Chern number in Fig.~\ref{fig:asym_con_dyn} (a), where we have set $M_1 = 
-0.9$, $M_2 = -0.45$, $t_x =1$ and $t_y = 2$.

We now consider the case when $\De_2 \ne 0$. The behavior at the momenta 
$\bk_\pm$ is the same as in the case $\De_2 = 0$, because at these points the 
$\De_2$ term vanishes, and the Hamiltonians for the two halves of the cycle 
continue to commute. However, this is not true at the momenta $\bk_{1,2}$ 
since the $\De_2$ term survives at these momenta. The time-evolution operator 
must therefore be written as 
\beq U_F (\bk_{1,2}) ~=~ U_1 U_2. \eeq
We would again like this to be equal to $\pm \mathbb{I}_4$ as in 
Eq.~\eqref{eq:Uid}. However, this would imply that $U_1$ and $U_2$ must be 
inverses of each other (possibly up to a sign) and hence must commute. This 
contradicts the fact that $U_1$ and $U_2$ do not commute at $\bk_{1,2}$. This 
contradiction implies that these two momenta cannot contribute to a change in 
any topological invariant. Indeed we see in Fig.~\ref{fig:mirror_dyn}
that the diagonal winding number jumps when $U_F (\bk) = \pm ~\mathbb{I}_4$
at $\bk = \bk_\pm$.

We can summarize the results presented here as follow. For $\De_2 = 0$, there 
are jumps in the Chern number whenever $U_F (\bk) = \pm \mathbb{I}_4$ at any 
of the four momenta $(0,0)$, $(\pi,\pi)$, $(0,\pi)$ and $(\pi,0)$. For any
value of $\De_2$, there are jumps in the diagonal winding number whenever 
$U_F (\bk) = \pm \mathbb{I}_4$ at any of the two momenta $(0,0)$ and 
$(\pi,\pi)$.

\section{Contributions to the Chern number from other momenta}
\label{app:k_points2}

We will now study if Eq.~\eqref{eq:Uid} can be satisfied at momenta 
which differ from the four time-reversal invariant points and have no
special symmetries. We will only consider the case $\De_2 = 0$ here. 
We would like to know whether the Floquet operator given
in Eq.~\eqref{eq:U_lat} can be equal to to $\pm \mathbb{I}_4$, when
$H_1 (\bk)$ and $H_2 (\bk)$ are given by Eq.~\eqref{eq:con_ham} or 
\eqref{eq:con_ham_asym} and $M = M_1$ and $M_2$ respectively. It is clear
that if $\bk$ is not at one of the four time-reversal invariant points,
$e^{-iH_1 (\bk)T/2}$ and $e^{-iH_2 (\bk)T/2}$ will generally not commute 
with each other. Hence Eq.~\eqref{eq:U_lat} can be equal to $\pm \mathbb{I}_4$
only if $e^{-iH_1 (\bk)T/2}$ and $e^{-iH_2 (\bk)T/2}$ are separately equal to 
$\pm \mathbb{I}_4$. Equation~\eqref{spec2} implies that this will happen if 
\bea && [ (M_1 ~+~ t_x \cos k_x ~+~ t_y \cos k_y))^2 \non \\
&& ~+~ \De_1^2 ~(\sin^2 k_x ~+~ \sin^2 k_y)]^{1/2} ~\frac{T}{2} ~=~ n_1 \pi, 
\non \\
&& [ (M_2 ~+~ t_x \cos k_x ~+~ t_y \cos k_y))^2 \non \\
&& ~+~ \De_1^2 ~(\sin^2 k_x ~+~ \sin^2 k_y)]^{1/2} ~\frac{T}{2} ~=~ n_2 \pi, 
\label{general_k} \eea
where $n_1$ and $n_2$ are positive integers. We now see that since there are 
two conditions to be satisfied, it may be possible to vary the two quantities 
$k_x$ and $k_y$ to satisfy Eqs.~\eqref{general_k}, provided that $T$ is
large enough. We also see that such solutions for $(k_x,k_y)$ will appear 
in groups of four since these equations remain unchanged if $k_x \to - k_x$
or $k_y \to - k_y$. Hence the Chern number may be expected to change by $\pm 
4$. This is precisely what we observe in Fig.~\ref{fig:asym_con_dyn} (a)
at $T \simeq 4.49$.

We can derive the precise value of $T$ where the Chern number jumps by 4 in 
Fig.~\ref{fig:asym_con_dyn} (a) as follows. Given that $M_1 = -0.9$, $M_2 = 
-0.45$, $\De_1 =1$, $t_x = 1$ and $t_y = 2$, we find numerically that there are
peaks in the Berry curvature near $k_x = \pm 1.44$ and $k_y = \pm 1.30$ when
$T$ is close to $4.49$. If we substitute these values of $k_x$, $k_y$ and $T$
in Eq.~\eqref{general_k}, we find that $n_1$ and $n_2$ are close to 1. Given 
this information, we can analytically derive the values of $k_x$, $k_y$ and 
$T$ which satisfy those equations with $n_1 = n_2 = 1$ exactly. We first 
observe that $(M_1 + t_x \cos k_x + t_y \cos k_y)^2$ must be equal to 
$(M_2 + t_x \cos k_x + t_y \cos k_y)^2$. Since $M_1 \ne M_2$, this implies that
\beq t_x \cos k_x ~+~ t_y \cos k_y ~=~ - ~\frac{1}{2} ~(M_1 ~+~ M_2). 
\label{cond1} \eeq
Next, the minimum value of $T$ where Eqs.~\eqref{general_k} will be satisfied
will correspond to the maximum value of
$(M_1 + t_x \cos k_x + t_y \cos k_y)^2 + \De_1^2 (\sin^2 k_x + \sin^2 k_y)$
as a function of $(k_x,k_y)$, subject to the condition in Eq.~\eqref{cond1}.
Solving this maximization problem, we obtain a second condition
\beq \frac{\cos k_x}{t_x} ~=~ \frac{\cos k_y}{t_y}. \label{cond2} \eeq
Using Eqs.~\eqref{cond1} and \eqref{cond2}, we find that
\bea \cos k_x &=& - ~\frac{(M_1 ~+~ M_2) ~t_x}{2 ~(t_x^2 ~+~ t_y^2)}, \non \\
\cos k_y &=& - ~\frac{(M_1 ~+~ M_2) ~t_y}{2 ~(t_x^2 ~+~ t_y^2)}, \eea
which gives four possible values of the momentum in the first Brillouin zone,
\beq k_x ~=~ \pm ~1.435 ~~~{\rm and}~~~ k_y ~=~ \pm 1.297. \label{kxky} \eeq
Note that the values of $(k_x,k_y)$ depend on the parameters $t_x, ~
t_y, ~M_1$ and $M_2$, unlike the time-reversal invariant momenta.
Substituting Eq.~\eqref{kxky} in Eqs.~\eqref{general_k} gives
\beq T ~=~ 4.489, \label{tc} \eeq
which agrees well with the value where the Chern number jumps by 4
in Fig.~\ref{fig:asym_con_dyn} (a).

The general solution to Eqs.~\eqref{general_k}
can be found as follows. Since $M_1 \ne M_2$, these equations imply that
\bea t_x \cos k_x ~+~ t_y \cos k_y &=& - ~\frac{1}{2} ~(M_1 ~+~ M_2) \non \\
&& +~ \frac{2 \pi^2 ~(n_1^2 ~-~ n_2^2)}{T^2 ~(M_1 ~-~ M_2)}. \label{cond3} \eea
Substituting this in one of the equations in Eqs.~\eqref{general_k},
and using the identity $\sin^2 k_x + \sin^2 k_y = 2 - \cos^2 k_x - \cos^2 k_y$
and Eq.~\eqref{cond3}, we obtain a quadratic equation for $\cos k_x$. This
generally gives two solutions for $\cos k_x$ and therefore for $\cos k_y$.
We therefore get four possible values of $k_x$ and two values of $k_y$ in each 
case, giving a total of eight different solutions for $(k_x,k_y)$. (In special 
cases, these eight solutions become degenerate and reduce to four solutions).

We find that as $T$ increases beyond the value given in Eq.~\eqref{tc}, 
there are more and more values of $(k_x,k_y)$ for which Eqs.~\eqref{general_k} 
are satisfied for various integer values of $n_1$ and $n_2$. This gives rise to
an increasing number of jumps in the Chern number.

In addition to the above momenta where the Floquet eigenvalues are 
exactly equal to $\pm 1$, we find that there are sometimes large numbers 
of momenta where the Floquet eigenvalues come very close to (but not exactly
equal to) $\pm 1$, and the quasienergy gaps become very small. As a result, 
the Berry curvature becomes very large near those points; this gives rise 
to sharp fluctuations in the Chern number, since this is obtained by 
integrating the Berry curvature over the Brillouin zone. We see examples 
of this in Fig.~\ref{fig:asym_con_dyn} (a) at $T=4.81$ and $4.91$. 
These fluctuations in the Chern number are numerical artifacts. Namely,
we have numerically studied the vicinity of the small quasienergy gaps 
using a very fine resolution in the momentum $(k_x,k_y)$, and we have
found that although the gaps are very small, they are not exactly zero.
The very small gaps lead to large fluctuations in the numerically calculated 
value of the Berry curvature and therefore of the Chern number.

\end{document}